\documentclass[letterpaper, onecolumn, 11pt, unpublished]{quantumarticle}
\pdfoutput=1

\usepackage[utf8]{inputenc}
\usepackage[english]{babel}
\usepackage[T1]{fontenc}
\usepackage[numbers, sort&compress]{natbib}

\usepackage{amsmath}
\usepackage{amssymb}
\usepackage{comment}
\usepackage{cancel} 
\usepackage{graphicx} 
\newcommand{\tr}{\operatorname{tr}} 
\newcommand{\Tr}{\operatorname{Tr}} 
\allowdisplaybreaks
\numberwithin{equation}{section}

\usepackage{hyperref}
\usepackage{dsfont} 
\usepackage{mathrsfs} 

\makeatletter
\DeclareFontFamily{OMX}{MnSymbolE}{}
\DeclareSymbolFont{MnLargeSymbols}{OMX}{MnSymbolE}{m}{n}
\SetSymbolFont{MnLargeSymbols}{bold}{OMX}{MnSymbolE}{b}{n}
\DeclareFontShape{OMX}{MnSymbolE}{m}{n}{
    <-6>  MnSymbolE5
   <6-7>  MnSymbolE6
   <7-8>  MnSymbolE7
   <8-9>  MnSymbolE8
   <9-10> MnSymbolE9
  <10-12> MnSymbolE10
  <12->   MnSymbolE12
}{}
\DeclareFontShape{OMX}{MnSymbolE}{b}{n}{
    <-6>  MnSymbolE-Bold5
   <6-7>  MnSymbolE-Bold6
   <7-8>  MnSymbolE-Bold7
   <8-9>  MnSymbolE-Bold8
   <9-10> MnSymbolE-Bold9
  <10-12> MnSymbolE-Bold10
  <12->   MnSymbolE-Bold12
}{}

\let\llangle\@undefined
\let\rrangle\@undefined
\DeclareMathDelimiter{\llangle}{\mathopen}%
                     {MnLargeSymbols}{'164}{MnLargeSymbols}{'164}
\DeclareMathDelimiter{\rrangle}{\mathclose}%
                     {MnLargeSymbols}{'171}{MnLargeSymbols}{'171}
\makeatother

\def\sbra#1{{\left\llangle #1 \right|}} 
\def\sket#1{{\left| #1 \right\rrangle}} 

\begin{document}

\title{Lifting Multiplicity in Randomized Benchmarking}

\author{Yale Fan}
\affiliation{Department of Physics, University of Idaho, Moscow, ID 83844, USA}
\email{yalefan@gmail.com}

\author{J. P. Marceaux}
\affiliation{IonQ Inc., Boulder, CO 80301, USA}
\email{jp.marceaux@oxionics.com}

\begingroup
\hbadness=10000 
\maketitle
\endgroup

\begin{abstract}
Randomized benchmarking comprises a suite of standard techniques for assessing the operational fidelity of quantum processors. These techniques can be reliably applied to groups with process matrix representations that are free of multiplicity. However, many groups that are naturally formed from native gates do have multiplicity, and it is difficult to draw robust conclusions from the resulting data. In this work, we present a new and intrinsic solution to the problem of multiplicity that completely eliminates the need to fit multi-exponential decays while still remaining robust to errors in state preparation and measurement. Our procedure, which we call lifted randomized benchmarking, generalizes key elements of both character and synthetic randomized benchmarking to benchmark a target group of quantum gates with respect to an enlarged group. We illustrate in numerical examples that our method can efficiently and reliably estimate the fidelity of two-qubit gates relevant to experimental trapped-ion systems. These techniques may find applications beyond randomized benchmarking to quantum tomography and learning, as well as the characterization of error-correcting codes.
\end{abstract}

\tableofcontents

\section{Introduction}

Understanding noise is the foundation of quantum error correction.  Randomized benchmarking (RB) \cite{RB-Emerson, RB-Dankert} has become a standard diagnostic of noise in quantum devices because it is simple, scalable, and robust to errors in state preparation and measurement (SPAM).  Standard RB samples gates uniformly at random from a unitary 2-design such as the multi-qubit Clifford group to extract a single average gate error rate from an exponential decay of survival probabilities \cite{CRB-Emerson, CRB-Knill, CRB-Magesan-1, CRB-Magesan-2}.  However, sampling from the Clifford group usually does not give an accurate picture of the noise affecting native gates, which motivates more general frameworks.  These general frameworks must surmount a challenging signal processing problem: for general gate groups, RB requires fitting multi-exponential decay curves.

Existing techniques such as character RB \cite{Helsen, Claes} and filtered RB \cite{FilteredRB-1, FilteredRB-2, Kong, Chen} partially mitigate this problem by constructing classical post-processing filters that isolate specific exponentially decaying signals.  However, these techniques can only reduce the problem to that of fitting single exponential decays in the case that the benchmarking group is ``multiplicity-free'' (we will explain this term shortly).  This is a significant restriction that does not capture many benchmarking groups of interest.  For the most general possible benchmarking groups, character RB is incapable of isolating noise contributions within certain ``multiplicity spaces,'' while filtered RB attempts to do so by directly fitting multi-exponential decays using complicated signal processing techniques that do not exploit the symmetry of the problem.  Benchmarking the most general ``non-multiplicity-free'' groups of quantum gates has remained a major problem in the theory of randomized benchmarking.

In this paper, we solve this problem by formulating a general procedure for randomized benchmarking of arbitrary gate groups that completely eliminates, as a matter of principle, the need to fit multi-exponential decays.  Our scheme allows us to reconstruct matrix exponential decays for non-multiplicity-free groups by fitting only single exponential decays.  These techniques could have many applications beyond randomized benchmarking.

We begin by giving both conceptual and technical overviews of our procedure.  In the rest of the paper, we review the necessary mathematical ingredients, describe the structure of what we call ``lifted RB'' in detail, discuss its practical implementation, and examine its applications to benchmarking two distinct groups of two-qubit gates.

\subsection{Conceptual Overview}

For non-multiplicity-free groups, character RB can only be used to extract \emph{matrix exponential decays}, which are sums of scalar exponential decays corresponding to individual eigenvalues.  We show how to extract the \emph{individual} eigenvalue decays in a SPAM-robust way by generalizing the character RB protocol, at the cost of enlarging the group of operations that one must implement.  To do so, we combine the strategy of enlarging the gate group with new synthetic RB techniques \cite{Fan:2024puh}.  Our approach can be viewed as an extension of character RB with three major differences: (1) we employ, as an intermediate step, a larger gate group instead of a smaller one (compare to \cite{Helsen2022matchgate}); (2) we use a generalization of the character formula for building projection operators; and (3) we synthesize operators (which generalize projectors) at both the beginning \emph{and} the end of the gate sequence.  In more detail, our methodology is as follows:
\begin{enumerate}
\item \textbf{Choose a larger group $F\supset G$} whose process matrix (superoperator) representation $\mathcal{F}$ has reduced multiplicity compared to $\mathcal{G}$.  In particular, choose $F$ such that the isotypic component of any given target irrep of $G$ inside $\mathcal{G}$ lies inside a single multiplicity-free irrep of $\mathcal{F}$.  Such a group $F$ always exists: for example, in the worst case, one may choose the entire unitary group on the system of interest.

This first step is the opposite of that taken in character RB, which involves choosing a subgroup $H\subset G$ and projecting onto a smaller irrep.  Rather than finding a subgroup of the benchmarking group, we embed the benchmarking group into a larger group that is ``sufficiently'' multiplicity-free.
\item \textbf{Design synthetic gates using physical gates from $F$ rather than $G$.}  This allows one to target signals from individual non-multiplicity-free irreps inside $\mathcal{G}$, which would be impossible using physical gates from $G$.  By implementing such synthetic gates (which generalize projection operators) at both the beginning and the end of each random gate sequence, one can perform full tomography on the symmetry-averaged error channel in a SPAM-robust way.\footnote{We emphasize that our variant of tomography employs synthetic gates rather than SPAM operations alone, which makes it robust to SPAM errors (unlike quantum process tomography).}

As a matter of terminology, a ``synthetic'' operation is a linear combination of physical gates, states, or measurements that is implemented through classical post-processing of measured data.  Synthetic RB \cite{Fan:2024puh} is a generalization of filtered RB using \emph{synthetic quantum circuits} that incorporate some combination of \emph{synthetic gates} and \emph{synthetic SPAM} operations.
\item \textbf{Execute RB using random gate sequences from $G$} of varying lengths and post-process the output data accordingly.  In effect, this procedure benchmarks $G$ with respect to the multiplicity-free structure of $F$.
\end{enumerate}
Altogether, our approach, which we refer to as ``lifted RB,'' could be thought of as a non-abelian variant of phase estimation that applies to matrices rather than to scalars.  Assuming a good choice of lifting group, our method can efficiently and unambiguously benchmark non-multiplicity-free groups, thereby overcoming a major limitation of prior RB protocols.

\subsection{Technical Overview}

We first set some notation.  For a general benchmarking group $G$, the conjugation representation on the space $\mathcal{L}(\mathscr{H})$ of operators on a given Hilbert space $\mathscr{H}$ (the ``superoperator representation'' $\mathcal{G}$) typically decomposes into many irreducible representations (irreps) indexed by $\lambda$.  If these irreps are multiplicity-free (in which case we say for brevity that $G$ itself is multiplicity-free), then RB survival probabilities take the form of a mixture of exponentials, $p_m = \sum_\lambda a_\lambda f_\lambda^m$, where $m$ is the gate sequence length and each decay parameter $f_\lambda$ probes noise projected onto a specific irrep of $\mathcal{G}$.  More generally, suppose that $G$ is not multiplicity-free, in which case the superoperator representation $\mathcal{G}$ of $G$ decomposes as $\mathcal{G}\cong \bigoplus_\lambda \left(\mathbb{C}^{n_\lambda}\otimes V_\lambda\right)$, where $n_\lambda$ is the multiplicity of irrep $\lambda$. This implies that the survival probability decomposes as $p_m = \sum_\lambda \Tr( A_\lambda M_\lambda^m)$, where $A_\lambda$ and $M_\lambda$ are square matrices of size $n_\lambda$. Character RB can be used to extract the matrix exponential decay associated with the isotypic component (i.e., the collection of all isomorphic copies) of any given irrep $\lambda$ in $\mathcal{G}$, namely $\Tr(A_\lambda M_\lambda^m)$.

We now fix $G$.  For any given target irrep $\phi_\lambda$ of $G$ that occurs (possibly with multiplicity) inside the superoperator representation $\mathcal{G}$, we choose a group $F\supset G$ whose superoperator representation $\mathcal{F}$ contains a multiplicity-free irrep $\phi_\Lambda$ that in turn contains the entire isotypic component of $\phi_\lambda$.  When we say that a representation $\Phi$ ``contains'' another representation $\phi$, we mean that the support (carrier space) $V_\phi$ of $\phi$ is a linear subspace of the support $V_\Phi$ of $\Phi$.  Therefore, in our case, the span of the subspaces $V_{\lambda^{(i)}}$ for all $i = 1, \ldots, n_\lambda$, where $n_\lambda$ is the multiplicity of $\phi_\lambda$, must be a subspace of $V_\Lambda$.  We can write this span as $\operatorname{span}(V_{\lambda^{(1)}}\cup \cdots\cup V_{\lambda^{(n_\lambda)}})$ or as $V_{\lambda^{(1)}} + \cdots + V_{\lambda^{(n_\lambda)}}$.

Using the formula
\begin{equation}
(\dim\phi_\Lambda)\mathbb{E}_{f\in F}[(\mathcal{F}|\mathcal{A})\mathcal{F}] = \mathcal{A},
\label{Fcompleteness}
\end{equation}
we can write any superoperator $\mathcal{A}$ that acts within the support of $\phi_\Lambda$ (and is zero everywhere else) as a linear combination of $F$ superoperator representation matrices.  Note that this formula holds for both finite and infinite (compact) groups, as $\mathbb{E}$ denotes an average.  Motivated by this formula, we run RB experiments using random gate sequences from $G$ that are bookended by random gates from $F$ to estimate the average weighted survival probabilities
\begin{equation}
p_m^\Lambda = (\dim\phi_\Lambda)^2\mathbb{E}_{\vec{g}\in G^m, f_L\in F, f_R\in F}[(\mathcal{F}_L|\mathcal{A}_L)(\mathcal{F}_R|\mathcal{A}_R)\sbra{E}\widetilde{\mathcal{F}_L\mathcal{G}_\text{inv}}\widetilde{\mathcal{G}}_m\cdots \widetilde{\mathcal{G}}_2\widetilde{\mathcal{G}_1\mathcal{F}_R}\sket{\rho}],
\end{equation}
where:
\begin{itemize}
\item $m$ is the sequence length (up to a constant offset),
\item $\sket{\rho}$ and $\sbra{E}$ are SPAM operations,
\item the initial $F$ gate and the initial $G$ gate, as well as the final $F$ gate and the final $G$ gate, are compiled into single gates (which are elements of $F$),
\item we are free to choose the operators $\mathcal{A}_L$ and $\mathcal{A}_R$, which both act within the support of the $F$ irrep $\phi_\Lambda$.
\end{itemize}
Tildes denote noisy gate implementations.  Let us assume, for simplicity, that all noisy gates are described by a single error channel $\mathcal{E}$.  Then this expression simplifies to
\begin{equation}
p_m^\Lambda = \sbra{E}\mathcal{E}\mathcal{A}_L\left(\sum_\lambda\sum_{i, i' = 1}^{n_\lambda} [M_\lambda(\mathcal{E})^m]_{ii'}Q_{\lambda, i, i'}\right)\mathcal{A}_R\sket{\rho},
\end{equation}
where:
\begin{itemize}
\item $M_\lambda(\mathcal{E})$ is the $n_\lambda\times n_\lambda$ matrix with entries
\begin{equation}
[M_\lambda(\mathcal{E})]_{ii'} = \frac{\tr(Q_{\lambda, i', i}\mathcal{E})}{\dim\phi_\lambda},
\end{equation}
\item $Q_{\lambda, i, i'}$ is the isomorphism between the $G$ irreps $\phi_{\lambda^{(i)}}$ and $\phi_{\lambda^{(i')}}$ given by
\begin{equation}
Q_{\lambda, i, i'} = \sum_{\ell=1}^{\dim\phi_\lambda} |\lambda^{(i)}, \ell\rrangle\llangle\lambda^{(i')}, \ell|,
\end{equation}
where $\{|\lambda^{(i)}, \ell\rangle \,|\, \ell = 1, \ldots, \dim\phi_\lambda\}$ is an orthonormal basis for the support of $\phi_{\lambda^{(i)}}$ such that the representation matrices for different copies $i$ of the same irrep $\lambda$ are identical.

As a matrix, the superoperator $Q_{\lambda, i, i'}$ has a single nonvanishing square block, which can be either a main-diagonal block or an off-diagonal block, and which takes the form of an identity matrix of size $\dim\phi_\lambda$.  This block occurs within the larger main-diagonal block of size $n_\lambda\dim\phi_\lambda$ that acts within the isotypic component of $\phi_\lambda$.
\end{itemize}
For example, if we choose $\mathcal{A}_L = Q_{\lambda, i_L, i_0}$ and $\mathcal{A}_R = Q_{\lambda, i_0', i_R}$ for some fixed $i_L, i_R$, then this expression further simplifies to
\begin{equation}
p_m^\Lambda = \sbra{E}\mathcal{E}Q_{\lambda, i_L, i_R}\sket{\rho}[M_\lambda(\mathcal{E})^m]_{i_0 i_0'}.
\end{equation}
By letting $i_0, i_0'$ range over $1, \ldots, n_\lambda$, we can estimate the full matrix $M_\lambda(\mathcal{E})^m$ for each $m$ and, in turn, $M_\lambda(\mathcal{E})$ itself, which we can then diagonalize to fit the individual eigenvalue decays.

For the above scheme to work, we must choose $F$ such that its superoperator representation contains a single multiplicity-free irrep that contains the entire isotypic component of a target irrep of $G$:
\begin{itemize}
\item $\phi_\Lambda$ must contain the entire isotypic component of $\phi_\lambda$ to ensure that we can synthesize all of the superoperators $Q_{\lambda, i, i'}$, which connect different copies $\phi_{\lambda^{(i)}}$ and $\phi_{\lambda^{(i')}}$.  This allows us to perform full tomography on the matrix $M_\lambda(\mathcal{E})$.  It is not enough for $\phi_\Lambda$ to contain, for instance, a single copy of $\phi_\lambda$.
\item $\phi_\Lambda$ must also be multiplicity-free to ensure that we can synthesize operators that act \emph{solely} within the target isotypic subspace.  This avoids contaminating the RB signal.  It is not enough to require that the isotypic component of $\phi_\lambda$ lie within the isotypic component of an irrep of $F$ that occurs with multiplicity.
\end{itemize}
See Appendix \ref{app:trivial} for a precise statement in the case that the target irrep is the trivial representation.  Note that if $G$ is a subgroup of a unitary 2-group such as the Clifford group, then such a group can play the role of $F$ for any target irrep $\phi_\lambda$.  Of course, the entire unitary group also satisfies the above criterion, but it is preferable from an experimental standpoint to sample from a finite unitary 2-design.

\section{Randomized Benchmarking with Multiplicity}

We begin by reviewing how randomized benchmarking works in the general setting of non-multiplicity-free benchmarking groups.  The relevant formulas from representation theory are derived in Appendix \ref{app:twirling}.

\subsection{Standard and Character RB}

A common subroutine at the core of many RB protocols is to sample uniformly at random from the set of gate sequences of length $m + 1$ that compile to a fixed ``ending gate'' $g_\text{end}$ for the purpose of estimating the average survival probability
\begin{equation}
p_m(g_\text{end}) = \mathbb{E}_{\vec{g}\in G^{m+1}, g_{m+1}\cdots g_1 = g_\text{end}}[\sbra{E}\widetilde{\mathcal{G}}_{m+1}\cdots \widetilde{\mathcal{G}}_1\sket{\rho}].
\end{equation}
In character (and, more generally, filtered) RB, we vary over $g_\text{end}$, which requires additional random sampling; in standard RB, we do not.

In standard RB, we set $g_\text{end} = e$.  That is, we run many circuits of $m$ uniformly randomly sampled gates $\vec{g}\equiv (g_1, \ldots, g_m)\in G^m$ followed by a global inversion gate $g_\text{inv} = (g_m\cdots g_1)^\dag$ to estimate the average survival probability
\begin{equation}
p_m = \mathbb{E}_{\vec{g}\in G^m}[\sbra{E}\widetilde{\mathcal{G}}_\text{inv}\widetilde{\mathcal{G}}_m\cdots \widetilde{\mathcal{G}}_1\sket{\rho}].
\end{equation}
Let us model noise by a single gate-independent error channel $\mathcal{E}$.  Regardless of whether the superoperator representation $\mathcal{G}$ of $G$ has multiplicity, we can write
\begin{align}
p_m &= \mathbb{E}_{\vec{g}\in G^m}[\sbra{E}(\mathcal{E}\mathcal{G}_1^{-1}\cdots \mathcal{G}_m^{-1})(\mathcal{E}\mathcal{G}_m)\cdots (\mathcal{E}\mathcal{G}_1)\sket{\rho}] \\
&= \mathbb{E}_{\vec{g}\in G^m}[\sbra{E}\mathcal{E}(\mathcal{G}_m^{-1}\mathcal{E}\mathcal{G}_m)\cdots (\mathcal{G}_1^{-1}\mathcal{E}\mathcal{G}_1)\sket{\rho}] \\
&= \sbra{E}\mathcal{E}\mathbb{E}_{g\in G}[\mathcal{G}^{-1}\mathcal{E}\mathcal{G}]^m\sket{\rho}, \label{avgsurvival}
\end{align}
where we have used the fact that the $G$-twirled error channel $\mathbb{E}_{g\in G}[\mathcal{G}^{-1}\mathcal{E}\mathcal{G}]$ commutes with any individual group element superoperator $\mathcal{G}$.  If the superoperator representation $\mathcal{G}$ decomposes into non-isomorphic irreps $\phi_\lambda$, then
\begin{equation}
\mathbb{E}_{g\in G}[\mathcal{G}^{-1}\mathcal{E}\mathcal{G}] = \sum_\lambda \frac{\Tr(\Pi_\lambda\mathcal{E})}{\Tr(\Pi_\lambda)}\Pi_\lambda,
\label{nomultiplicitytwirl}
\end{equation}
where $\Pi_\lambda$ is the projector onto the support of $\phi_\lambda$.  In this case, \eqref{avgsurvival} becomes
\begin{equation}
p_m = \sum_\lambda \sbra{E}\mathcal{E}\Pi_\lambda\sket{\rho}\left[\frac{\Tr(\Pi_\lambda\mathcal{E})}{\Tr(\Pi_\lambda)}\right]^m.
\end{equation}
More generally, if the representation $\mathcal{G}$ has multiplicity, then---in a basis where the representation matrices of isomorphic irreps are identical---the $G$-twirl takes the form
\begin{equation}
\mathbb{E}_{g\in G}[\mathcal{G}^{-1}\mathcal{E}\mathcal{G}] = \bigoplus_\lambda M_\lambda(\mathcal{E})\otimes \mathds{1}_\lambda,
\label{multiplicitytwirl}
\end{equation}
where $\mathds{1}_\lambda$ is the $(\dim\phi_\lambda)$-dimensional identity matrix, $M_\lambda(\mathcal{E})$ is an $n_\lambda\times n_\lambda$ matrix with entries
\begin{equation}
[M_\lambda(\mathcal{E})]_{ii'} = \frac{\tr(Q_{\lambda, i', i}\mathcal{E})}{\dim\phi_\lambda},
\end{equation}
and $Q_{\lambda, i, i'}$ is the isomorphism from irrep $\phi_{\lambda^{(i')}}$ to irrep $\phi_{\lambda^{(i)}}$ that preserves the matrix elements of the representation matrices.  In particular, we have $\Pi_{\lambda^{(i)}} = Q_{\lambda, i, i}$, while the projector onto a single isotypic component is given by $\Pi_\lambda = \sum_{i = 1}^{n_\lambda} \Pi_{\lambda^{(i)}}$.  Note that although the group element superoperators within each isotypic component decompose into separate blocks for each irrep, the twirled error channels do not.  Since
\begin{equation}
\mathbb{E}_{g\in G}[\mathcal{G}^{-1}\mathcal{E}\mathcal{G}]^m = \bigoplus_\lambda M_\lambda(\mathcal{E})^m\otimes \mathds{1}_\lambda,
\end{equation}
matrix exponential decays are governed by the eigenvalues of $M_\lambda(\mathcal{E})$.  In a different basis, the twirl $\mathbb{E}_{g\in G}[\mathcal{G}^{-1}\mathcal{E}\mathcal{G}]$ would be conjugated by a unitary matrix, which would only affect the pre\-factors in the matrix exponential decays.  As long as $M_\lambda(\mathcal{E})$ is diagonalizable, we can write\footnote{In particular, if $M_\lambda(\mathcal{E})$ is Hermitian, then its eigenvalues are real and one can write its spectral decomposition in terms of an orthonormal basis of eigenvectors.  If $M_\lambda(\mathcal{E})$ is real, then its complex eigenvalues come in complex conjugate pairs.  In general, only the spectrum of the full twirled noise channel need be invariant under complex conjugation because channels, as positive linear maps, preserve hermiticity.  Complex eigenvalues lead to oscillations in the RB signal.}
\begin{equation}
M_\lambda(\mathcal{E}) = V_\lambda(\mathcal{E})\Gamma_\lambda(\mathcal{E})V_\lambda(\mathcal{E})^{-1},
\label{diagonalization}
\end{equation}
where the columns of $V_\lambda(\mathcal{E})$ are the eigenvectors of $M_\lambda(\mathcal{E})$ and
\begin{equation}
\Gamma_\lambda(\mathcal{E}) = \operatorname{diag}(\gamma_{\lambda, 1}(\mathcal{E}), \ldots, \gamma_{\lambda, n_\lambda}(\mathcal{E})).
\end{equation}
(Equivalently, one can write the spectral decomposition of any diagonalizable matrix in terms of biorthogonal left and right eigenvectors, as in \cite{Claes}.  For $M_\lambda(\mathcal{E})$, the right eigenvectors are the columns of $V_\lambda(\mathcal{E})$, and the left eigenvectors are the columns of $(V_\lambda(\mathcal{E})^{-1})^\dag$.) Hence
\begin{equation}
\mathbb{E}_{g\in G}[\mathcal{G}^{-1}\mathcal{E}\mathcal{G}]^m = \left(\bigoplus_\mu V_\mu(\mathcal{E})\otimes \mathds{1}_\mu\right)\left(\bigoplus_\lambda \Gamma_\lambda(\mathcal{E})^m\otimes \mathds{1}_\lambda\right)\left(\bigoplus_\nu V_\nu(\mathcal{E})^{-1}\otimes \mathds{1}_\nu\right),
\end{equation}
where
\begin{equation}
\bigoplus_\lambda \Gamma_\lambda(\mathcal{E})^m\otimes \mathds{1}_\lambda = \sum_\lambda \sum_{i=1}^{n_\lambda} \gamma_{\lambda, i}(\mathcal{E})^m\Pi_{\lambda^{(i)}}.
\label{diagonalpowers}
\end{equation}
Substituting into \eqref{avgsurvival}, we obtain
\begin{align}
p_m &= \sum_\lambda \sum_{i=1}^{n_\lambda} \sbra{E}\mathcal{E}\left(\bigoplus_\mu V_\mu(\mathcal{E})\otimes \mathds{1}_\mu\right)\Pi_{\lambda^{(i)}}\left(\bigoplus_\nu V_\nu(\mathcal{E})^{-1}\otimes \mathds{1}_\nu\right)\sket{\rho}\gamma_{\lambda, i}(\mathcal{E})^m \\
&= \sum_\lambda \sum_{i=1}^{n_\lambda} \sbra{E}\mathcal{E}\left[\left(V_\lambda(\mathcal{E})(E_{n_\lambda})_{ii}V_\lambda(\mathcal{E})^{-1}\otimes \mathds{1}_\lambda\right)\oplus \bigoplus_{\xi\neq \lambda} \mathbf{0}_{n_\xi\dim\phi_\xi}\right]\sket{\rho}\gamma_{\lambda, i}(\mathcal{E})^m,
\end{align}
where $(E_D)_{ii}$ is the $D\times D$ matrix unit with nonzero $(i, i)$ entry and $\mathbf{0}_D$ is the $D\times D$ zero matrix.\footnote{In terms of $(E_D)_{ii}$, we can also write $Q_{\lambda, i, i'} = ((E_{n_\lambda})_{ii'}\otimes \mathds{1}_\lambda)\oplus \bigoplus_{\xi\neq \lambda} \mathbf{0}_{n_\xi\dim\phi_\xi}$.}  This formula makes clear that the coefficients depend on the support of the SPAM operations within each isotypic component and not only within individual irreps.

In character RB, we choose a subgroup $H\subset G$ (the \emph{character group}) whose superoperator representation $\mathcal{H}$ contains an irrep $\phi_{\lambda'}$ whose isotypic component has support inside the isotypic component of a target irrep $\phi_\lambda$ of $G$.  We then set $g_\text{end} = h\in H$ to estimate the average \emph{weighted} survival probability
\begin{equation}
p_m^{\lambda'} = (\dim\phi_{\lambda'})\mathbb{E}_{\vec{g}\in G^m, h\in H}[\chi_{\lambda'}(h)^\ast\sbra{E}\widetilde{\mathcal{G}}_\text{inv}\widetilde{\mathcal{G}}_m\cdots \widetilde{\mathcal{G}}_2\widetilde{\mathcal{G}_1\mathcal{H}}\sket{\rho}].
\end{equation}
Using that
\begin{equation}
(\dim\phi_\lambda)\mathbb{E}_{g\in G}[\chi_\lambda(g)^\ast\mathcal{G}] = \Pi_\lambda
\label{projectorformula}
\end{equation}
where $\chi_\lambda$ is the character of the irrep $\phi_\lambda$ and $\Pi_\lambda$ is an isotypic projector, we can write
\begin{align}
p_m^{\lambda'} &= (\dim\phi_{\lambda'})\mathbb{E}_{\vec{g}\in G^m, h\in H}[\chi_{\lambda'}(h)^\ast\sbra{E}(\mathcal{E}\mathcal{G}_1^{-1}\cdots \mathcal{G}_m^{-1})(\mathcal{E}\mathcal{G}_m)\cdots (\mathcal{E}\mathcal{G}_2)(\mathcal{E}\mathcal{G}_1\mathcal{H})\sket{\rho}] \\
&= \sbra{E}\mathcal{E}\mathbb{E}_{g\in G}[\mathcal{G}^{-1}\mathcal{E}\mathcal{G}]^m\mathbb{E}_{h\in H}[(\dim\phi_{\lambda'})\chi_{\lambda'}(h)^\ast\mathcal{H}]\sket{\rho} \\
&= \sbra{E}\mathcal{E}\mathbb{E}_{g\in G}[\mathcal{G}^{-1}\mathcal{E}\mathcal{G}]^m\Pi_{\lambda'}\sket{\rho}. \label{avgweightedsurvival}
\end{align}
If the representation $\mathcal{G}$ is multiplicity-free, then \eqref{nomultiplicitytwirl} implies that \eqref{avgweightedsurvival} simplifies to
\begin{equation}
p_m^{\lambda'} = \sbra{E}\mathcal{E}\Pi_{\lambda'}\sket{\rho}\left[\frac{\Tr(\Pi_\lambda\mathcal{E})}{\Tr(\Pi_\lambda)}\right]^m.
\end{equation}
If the representation $\mathcal{G}$ has multiplicity, then \eqref{multiplicitytwirl} yields
\begin{align}
p_m^{\lambda'} &= \sbra{E}\mathcal{E}\left(\bigoplus_\lambda M_\lambda(\mathcal{E})^m\otimes \mathds{1}_\lambda\right)\Pi_{\lambda'}\sket{\rho} \\
&= \sbra{E}\mathcal{E}\left(\bigoplus_\mu V_\mu(\mathcal{E})\otimes \mathds{1}_\mu\right)\left(\sum_\lambda \sum_{i=1}^{n_\lambda} \gamma_{\lambda, i}(\mathcal{E})^m\Pi_{\lambda^{(i)}}\right)\left(\bigoplus_\nu V_\nu(\mathcal{E})^{-1}\otimes \mathds{1}_\nu\right)\Pi_{\lambda'}\sket{\rho},
\end{align}
where we have diagonalized $M_\lambda(\mathcal{E})$ as before.  Since the isotypic component of $\lambda'$ lies within that of the target irrep $\lambda$, we isolate the decays corresponding to $\lambda$:
\begin{align}
p_m^{\lambda'} &= \sum_{i=1}^{n_\lambda} \sbra{E}\mathcal{E}\left(\bigoplus_\mu V_\mu(\mathcal{E})\otimes \mathds{1}_\mu\right)\Pi_{\lambda^{(i)}}\left(\bigoplus_\nu V_\nu(\mathcal{E})^{-1}\otimes \mathds{1}_\nu\right)\Pi_{\lambda'}\sket{\rho}\gamma_{\lambda, i}(\mathcal{E})^m \\
&= \sum_{i=1}^{n_\lambda} \sbra{E}\mathcal{E}\left[\left(V_\lambda(\mathcal{E})(E_{n_\lambda})_{ii}V_\lambda(\mathcal{E})^{-1}\otimes \mathds{1}_\lambda\right)\oplus \bigoplus_{\xi\neq \lambda} \mathbf{0}_{n_\xi\dim\phi_\xi}\right]\Pi_{\lambda'}\sket{\rho}\gamma_{\lambda, i}(\mathcal{E})^m.
\end{align}
This is a matrix exponential decay.  Note that one can choose different character groups for different target irreps (indeed, it may not be possible to choose a single character group that works for all irreps of the benchmarking group).

\subsection{Synthetic RB and Beyond}

Synthetic RB refers to a collection of techniques that generalize both the synthetic gates of character RB and the synthetic measurements of filtered RB \cite{Fan:2024puh}.

In particular, synthetic-SPAM RB unifies and generalizes certain tricks scattered throughout the RB literature (which require no representation-theoretic weighting of survival probabilities over different gate sequences) for distilling single exponential decays from linear combinations of multi-exponential decays.  Even without synthetic gates, one can use synthetic SPAM to improve the quality of RB signals with little to no additional effort on the experimental side.  In an experimental context where SPAM operations are reliable, synthetic-SPAM RB can isolate individual decays far more efficiently than character RB, even in cases without multiplicity.

In synthetic RB without multiplicity, the primary tool for building synthetic gates is the following completeness relation, which generalizes the isotypic projector formula used in character RB.  Any superoperator $\mathcal{A}$ that acts within a multiplicity-free irrep $\lambda$ occurring in the superoperator representation $\mathcal{G}$ of $G$ (meaning $\Pi_\lambda\mathcal{A}\Pi_\lambda = \mathcal{A}$) can be expressed as
\begin{equation}
(\dim\phi_\lambda)\mathbb{E}_{g\in G}[(\mathcal{G}|\mathcal{A})\mathcal{G}] = \mathcal{A},
\label{completeness}
\end{equation}
where $(\mathcal{A}|\mathcal{B}) = \Tr(\mathcal{A}^\dag\mathcal{B})$.  If $\mathcal{A} = \Pi_\lambda$, then \eqref{completeness} reduces to \eqref{projectorformula} since $\chi_\lambda(g) = \Tr(\Pi_\lambda\mathcal{G})$.

The synthetic technique faces the fundamental limitation that if physical gates are drawn from the benchmarking group $G$, then the corresponding synthetic gates can only project onto isotypic components of a given irrep rather than individual irreps themselves.  Indeed, if a superoperator $\mathcal{A}$ acts within the isotypic component of an irrep $\lambda$ that occurs with multiplicity $n_\lambda$ in the superoperator representation $\mathcal{G}$ (meaning $\Pi_\lambda\mathcal{A}\Pi_\lambda = \mathcal{A}$ where $\Pi_\lambda$ is understood as the isotypic projector), then \eqref{completeness} generalizes to
\begin{equation}
(\dim\phi_\lambda)\mathbb{E}_{g\in G}[(\mathcal{G}|\mathcal{A})\mathcal{G}] = \sum_{i, i' = 1}^{n_\lambda} Q_{\lambda, i, i'}\mathcal{A}Q_{\lambda, i', i}.
\label{completenessgen}
\end{equation}
We derive this formula in Appendix \ref{app:completeness}.  Even if we go beyond the group averaging approach by using fixed frames, any approach that uses group representation matrices is restricted by linearity to synthesizing operators that act identically within copies (or, by changes of basis, up to conjugation by unitaries that differ across copies).

Note that nothing prevents us from synthesizing SPAM operations within individual copies, but we need to be able to synthesize operators within those copies to ensure SPAM-robustness.  Therefore, while the ``multiplicity problem'' can in principle be solved by synthetic SPAM alone, a SPAM-robust protocol requires a new approach to synthetic gates.

\section{Lifted Randomized Benchmarking}

\subsection{Enlarging the Benchmarking Group}

We have seen that using physical gates from subgroups of the benchmarking group $G$ limits one to constructing projectors onto isotypic components of irreps of $G$.  To circumvent this obstacle, we use a larger group to synthesize (generalized) projection operators, but we use the original group for twirling.  This strategy allows us to benchmark only the errors associated with the original group.

To be precise, we choose a group $F\supset G$ whose superoperator representation $\mathcal{F}$ contains a multiplicity-free irrep $\phi_\Lambda$ that in turn contains the isotypic component of a target irrep $\phi_\lambda$ of $G$ occurring inside the superoperator representation $\mathcal{G}$.  Using $F$ superoperator representation matrices, we may synthesize superoperators $\mathcal{A}$ that act within any copy $\phi_{\lambda^{(i)}}$ (and therefore also within $\phi_\Lambda$) via the formula \eqref{Fcompleteness}, which was previously exploited in \cite{Fan:2024puh}. (To ease terminology, we sometimes refer to the support, or the carrier space, of an irrep as the irrep itself.)

We introduce an additional novel ingredient relative to character RB, which is that of implementing synthetic gates at both the beginning and the end of the random gate sequence.  To motivate this ``two-sided'' procedure, consider (for illustration) imitating the ``one-sided'' procedure of character RB to estimate the average weighted survival probability
\begin{equation}
\tilde{p}_m^\Lambda = (\dim\phi_\Lambda)\mathbb{E}_{\vec{g}\in G^m, f\in F}[(\mathcal{F}|\mathcal{A})\sbra{E}\widetilde{\mathcal{G}}_\text{inv}\widetilde{\mathcal{G}}_m\cdots \widetilde{\mathcal{G}}_2\widetilde{\mathcal{G}_1\mathcal{F}}\sket{\rho}].
\end{equation}
Suppose that all noisy $G$ superoperators are associated with the error channel $\mathcal{E}$, while noisy $F$ superoperators have a generally different error channel $\mathcal{E}'$.\footnote{Since some $F$ superoperators are also $G$ superoperators, $\mathcal{E}'$ cannot always be distinct from $\mathcal{E}$, but let us suppose that this is so.  This distinction is not essential: we could model the error with a single channel $\mathcal{E}$.}  In effect, we distinguish between the gates that we wish to benchmark and the totality of gates that we can implement, which may have different noise channels.  Then
\begin{align}
\tilde{p}_m^\Lambda &= (\dim\phi_\Lambda)\mathbb{E}_{\vec{g}\in G^m, f\in F}[(\mathcal{F}|\mathcal{A})\sbra{E}(\mathcal{E}\mathcal{G}_1^{-1}\cdots \mathcal{G}_m^{-1})(\mathcal{E}\mathcal{G}_m)\cdots (\mathcal{E}\mathcal{G}_2)(\mathcal{E}'\mathcal{G}_1\mathcal{F})\sket{\rho}] \\
&= \sbra{E}\mathcal{E}\mathbb{E}_{g\in G}[\mathcal{G}^{-1}\mathcal{E}\mathcal{G}]^{m-1}\mathbb{E}_{g\in G}[\mathcal{G}^{-1}\mathcal{E}'\mathcal{G}]\mathbb{E}_{f\in F}[(\dim\phi_\Lambda)(\mathcal{F}|\mathcal{A})\mathcal{F}]\sket{\rho}.
\end{align}
In the multiplicity-free case, we may choose $\mathcal{A}$ to act within the irrep $\lambda$, in which case \eqref{nomultiplicitytwirl} gives
\begin{equation}
\tilde{p}_m^\Lambda = \sbra{E}\mathcal{E}\mathcal{A}\sket{\rho}\left[\frac{\Tr(\Pi_\lambda\mathcal{E}')}{\Tr(\Pi_\lambda)}\right]\left[\frac{\Tr(\Pi_\lambda\mathcal{E})}{\Tr(\Pi_\lambda)}\right]^{m-1},
\end{equation}
which is a single exponentially decaying function of $m$.  With multiplicity, \eqref{multiplicitytwirl} gives
\begin{equation}
\tilde{p}_m^\Lambda = \sbra{E}\mathcal{E}\left(\bigoplus_\lambda M_\lambda(\mathcal{E})^{m-1}M_\lambda(\mathcal{E}')\otimes \mathds{1}_\lambda\right)\mathcal{A}\sket{\rho}.
\label{withmultiplicity}
\end{equation}
The matrix exponential decay is governed by the eigenvalues of $M_\lambda(\mathcal{E})$; everything else in this expression contributes a prefactor.  Note that if we had not compiled $f$ and $g_1$ into a single gate, then we would have obtained
\begin{align}
\tilde{p}_m^\Lambda &\stackrel{!}{=} (\dim\phi_\Lambda)\mathbb{E}_{\vec{g}\in G^m, f\in F}[(\mathcal{F}|\mathcal{A})\sbra{E}\widetilde{\mathcal{G}}_\text{inv}\widetilde{\mathcal{G}}_m\cdots \widetilde{\mathcal{G}_1}\widetilde{\mathcal{F}}\sket{\rho}] \\
&= (\dim\phi_\Lambda)\mathbb{E}_{\vec{g}\in G^m, f\in F}[(\mathcal{F}|\mathcal{A})\sbra{E}(\mathcal{E}\mathcal{G}_1^{-1}\cdots \mathcal{G}_m^{-1})(\mathcal{E}\mathcal{G}_m)\cdots (\mathcal{E}\mathcal{G}_1)(\mathcal{E}'\mathcal{F})\sket{\rho}] \\
&= \sbra{E}\mathcal{E}\mathbb{E}_{g\in G}[\mathcal{G}^{-1}\mathcal{E}\mathcal{G}]^m\mathcal{E}'\mathbb{E}_{f\in F}[(\dim\phi_\Lambda)(\mathcal{F}|\mathcal{A})\mathcal{F}]\sket{\rho},
\end{align}
which simplifies to
\begin{equation}
\tilde{p}_m^\Lambda = \sbra{E}\mathcal{E}\left(\bigoplus_\lambda M_\lambda(\mathcal{E})^m\otimes \mathds{1}_\lambda\right)\mathcal{E}'\mathcal{A}\sket{\rho}.
\end{equation}
This is not ideal because the presence of the untwirled error channel $\mathcal{E}'$ inhibits our ability to probe $M_\lambda(\mathcal{E})$ by choosing $\mathcal{A}$.  Returning to \eqref{withmultiplicity}, diagonalizing $M_\lambda(\mathcal{E})$ as in \eqref{diagonalization} and using \eqref{diagonalpowers} gives
\begin{align}
\tilde{p}_m^\Lambda &= \sbra{E}\mathcal{E}\left(\bigoplus_\mu V_\mu(\mathcal{E})\otimes \mathds{1}_\mu\right)\left(\bigoplus_\lambda \Gamma_\lambda(\mathcal{E})^{m-1}\otimes \mathds{1}_\lambda\right)\left(\bigoplus_\nu V_\nu(\mathcal{E})^{-1}M_\nu(\mathcal{E}')\otimes \mathds{1}_\nu\right)\mathcal{A}\sket{\rho} \nonumber \\
&= \sum_\lambda \sum_{i=1}^{n_\lambda} \sbra{E}\mathcal{E}\left(\bigoplus_\mu V_\mu(\mathcal{E})\otimes \mathds{1}_\mu\right)\Pi_{\lambda^{(i)}}\left(\bigoplus_\nu V_\nu(\mathcal{E})^{-1}M_\nu(\mathcal{E}')\otimes \mathds{1}_\nu\right)\mathcal{A}\sket{\rho}\gamma_{\lambda, i}(\mathcal{E})^{m-1} \nonumber \\
&= \sum_\lambda \sum_{i=1}^{n_\lambda} \sbra{E}\mathcal{E}\left[\left(V_\lambda(\mathcal{E})(E_{n_\lambda})_{ii}V_\lambda(\mathcal{E})^{-1}M_\lambda(\mathcal{E}')\otimes \mathds{1}_\lambda\right)\oplus \bigoplus_{\xi\neq \lambda} \mathbf{0}_{n_\xi\dim\phi_\xi}\right]\mathcal{A}\sket{\rho}\gamma_{\lambda, i}(\mathcal{E})^{m-1},
\end{align}
where we have used $\Pi_{\lambda^{(i)}} = \left((E_{n_\lambda})_{ii}\otimes \mathds{1}_\lambda\right)\oplus \bigoplus_{\xi\neq \lambda} \mathbf{0}_{n_\xi\dim\phi_\xi}$.  To say that $\mathcal{A}$ acts within $\phi_{\lambda^{(i)}}$ means that
\begin{equation}
\mathcal{A} = \left((E_{n_\lambda})_{ii}\otimes A_\lambda\right)\oplus \bigoplus_{\xi\neq \lambda} \mathbf{0}_{n_\xi\dim\phi_\xi}
\label{notgoodenough}
\end{equation}
for some arbitrary $(\dim\phi_\lambda)$-dimensional matrix $A_\lambda$.  However, choosing $\mathcal{A}$ to take this form is not quite good enough\footnote{If both $M_\lambda(\mathcal{E})$ and $M_\lambda(\mathcal{E}')$ were diagonal, then synthesizing an operator of the form \eqref{notgoodenough} would suffice to extract the entries.  This would be the case if the error channels $\mathcal{E}$ and $\mathcal{E}'$ were block diagonal with respect to the irrep structure of $\mathcal{G}$ (not just with respect to the isotypic decomposition, but with respect to the individual irreps).  Also, in the limit of small noise, $\mathcal{E}$ and $\mathcal{E}'$ are close to the identity channel, so $M_\lambda(\mathcal{E})$ and $M_\lambda(\mathcal{E}')$ are approximately diagonal.  However, the diagonal entries are not good approximations to the eigenvalues even in this limit because the off-diagonal entries are of the same order in the perturbation as the diagonal entries.  One would need to use degenerate perturbation theory.} to extract the signal from $\phi_{\lambda^{(i)}}$.  Extracting a single decay actually requires synthesizing an operator of the form
\begin{equation}
\mathcal{A} = \left(M_\lambda(\mathcal{E}')^{-1}V_\lambda(\mathcal{E})(E_{n_\lambda})_{ii}V_\lambda(\mathcal{E})^{-1}\otimes A_\lambda\right)\oplus \bigoplus_{\xi\neq \lambda} \mathbf{0}_{n_\xi\dim\phi_\xi},
\end{equation}
in which case we would have
\begin{equation}
\tilde{p}_m^\Lambda = \sbra{E}\mathcal{E}\left[\left(V_\lambda(\mathcal{E})(E_{n_\lambda})_{ii}V_\lambda(\mathcal{E})^{-1}\otimes A_\lambda\right)\oplus \bigoplus_{\xi\neq \lambda} \mathbf{0}_{n_\xi\dim\phi_\xi}\right]\sket{\rho}\gamma_{\lambda, i}(\mathcal{E})^{m-1}.
\end{equation}
In practice, of course, $V_\lambda(\mathcal{E})$ and $M_\lambda(\mathcal{E}')$ are unknown.  To surmount this problem, we instead perform full tomography on the matrix $M_\lambda(\mathcal{E})^{m-1}M_\lambda(\mathcal{E}')$ in a SPAM-robust way.  We do so by synthesizing operators on both the left \emph{and} the right to isolate individual matrix elements.  We then fit a single matrix exponential function to estimate $M_\lambda(\mathcal{E})$.

Our strategy is thus to synthesize two operators $\mathcal{A}_L$ and $\mathcal{A}_R$ to estimate the quantity
\begin{equation}
p_m^\Lambda = \sbra{E}\mathcal{O}_m^\Lambda\sket{\rho}
\end{equation}
where
\begin{align}
\mathcal{O}_m^\Lambda &= (\dim\phi_\Lambda)^2\underset{\substack{\vec{g}\in G^m \\ f_L\in F \\ f_R\in F}}{\mathbb{E}}[(\mathcal{F}_L|\mathcal{A}_L)(\mathcal{F}_R|\mathcal{A}_R)\widetilde{\mathcal{F}_L\mathcal{G}_\text{inv}}\widetilde{\mathcal{G}}_m\cdots \widetilde{\mathcal{G}}_2\widetilde{\mathcal{G}_1\mathcal{F}_R}] \\
&= (\dim\phi_\Lambda)^2\underset{\substack{\vec{g}\in G^m \\ f_L\in F \\ f_R\in F}}{\mathbb{E}}[(\mathcal{F}_L|\mathcal{A}_L)(\mathcal{F}_R|\mathcal{A}_R)(\mathcal{E}'\mathcal{F}_L\mathcal{G}_1^{-1}\cdots \mathcal{G}_m^{-1})(\mathcal{E}\mathcal{G}_m)\cdots (\mathcal{E}\mathcal{G}_2)(\mathcal{E}'\mathcal{G}_1\mathcal{F}_R)] \\
&= \mathcal{E}'\underset{f\in F}{\mathbb{E}}[(\dim\phi_\Lambda)(\mathcal{F}|\mathcal{A}_L)\mathcal{F}]\underset{g\in G}{\mathbb{E}}[\mathcal{G}^{-1}\mathcal{E}\mathcal{G}]^{m-1}\underset{g\in G}{\mathbb{E}}[\mathcal{G}^{-1}\mathcal{E}'\mathcal{G}]\underset{f\in F}{\mathbb{E}}[(\dim\phi_\Lambda)(\mathcal{F}|\mathcal{A}_R)\mathcal{F}],
\end{align}
which simplifies to
\begin{align}
p_m^\Lambda &= \sbra{E}\mathcal{E}'\mathcal{A}_L\left(\bigoplus_\lambda M_\lambda(\mathcal{E})^{m-1}M_\lambda(\mathcal{E}')\otimes \mathds{1}_\lambda\right)\mathcal{A}_R\sket{\rho} \\
&= \sbra{E}\mathcal{E}'\mathcal{A}_L\left(\sum_\lambda\sum_{i, i' = 1}^{n_\lambda} [M_\lambda(\mathcal{E})^{m-1}M_\lambda(\mathcal{E}')]_{ii'}Q_{\lambda, i, i'}\right)\mathcal{A}_R\sket{\rho}.
\end{align}
Note that $Q_{\lambda, i, i'}Q_{\lambda, j, j'} = \delta_{i'j}Q_{\lambda, i, j'}$.  To target the entry $[M_\lambda(\mathcal{E})^{m-1}M_\lambda(\mathcal{E}')]_{i_0 i_0'}$, we can therefore synthesize
\begin{equation}
\mathcal{A}_L = Q_{\lambda, i_L, i_0}, \qquad \mathcal{A}_R = Q_{\lambda, i_0', i_R}
\end{equation}
for some fixed indices $i_L, i_R$ (this is the step that makes use of the crucial assumption that $\phi_\Lambda$ contains \emph{the entire isotypic component of} $\phi_\lambda$).  This gives
\begin{equation}
p_m^\Lambda = \sbra{E}\mathcal{E}'Q_{\lambda, i_L, i_R}\sket{\rho}[M_\lambda(\mathcal{E})^{m-1}M_\lambda(\mathcal{E}')]_{i_0 i_0'}.
\end{equation}
Finally, we let $i_0, i_0'$ range over $1, \ldots, n_\lambda$.  For the sake of sample efficiency \cite{Fan:2024puh}, we could also synthesize operators of reduced rank, such as
\begin{equation}
\mathcal{A}_L = |\lambda^{(i_L)}, \ell_0\rrangle\llangle\lambda^{(i_0)}, \ell_0|, \qquad \mathcal{A}_R = |\lambda^{(i_0')}, \ell_0\rrangle\llangle\lambda^{(i_R)}, \ell_0|
\end{equation}
for some fixed $\ell_0$, in which case we would have
\begin{equation}
p_m^\Lambda = \llangle E|\mathcal{E}'|\lambda^{(i_L)}, \ell_0\rrangle\llangle\lambda^{(i_R)}, \ell_0|\rho\rrangle[M_\lambda(\mathcal{E})^{m-1}M_\lambda(\mathcal{E}')]_{i_0 i_0'}.
\end{equation}
We could then increase the overlaps (SPAM coefficients) using synthetic SPAM.

If $\mathcal{E}' = \mathcal{E}$, then we could estimate and diagonalize the matrix $M_\lambda(\mathcal{E})^m$ for each $m$ and then fit the individual eigenvalue decays.  If $\mathcal{E}'\neq \mathcal{E}$, then we cannot do this because diagonalizing the matrix $M_\lambda(\mathcal{E})^{m-1}M_\lambda(\mathcal{E}')$ would not yield (powers of) the eigenvalues of $M_\lambda(\mathcal{E})$.  Instead, we can multiply the estimate at sequence length $m$ by the inverse of the estimate at sequence length $m - 1$ (in either order) to obtain either
\begin{equation}
M_\lambda(\mathcal{E})^m M_\lambda(\mathcal{E}')[M_\lambda(\mathcal{E})^{m-1}M_\lambda(\mathcal{E}')]^{-1} = M_\lambda(\mathcal{E})
\end{equation}
or
\begin{equation}
[M_\lambda(\mathcal{E})^{m-1}M_\lambda(\mathcal{E}')]^{-1}M_\lambda(\mathcal{E})^m M_\lambda(\mathcal{E}') = M_\lambda(\mathcal{E}')^{-1}M_\lambda(\mathcal{E})M_\lambda(\mathcal{E}').
\end{equation}
Either way, diagonalizing the resulting matrix yields the eigenvalues of $M_\lambda(\mathcal{E})$.

\subsubsection{Synthesizing SPAM-Adapted Lifting Operators} \label{subsec:spam_lifting}

Synthesizing $Q_{\lambda, i, i'}$ directly poses some practical limitations. In particular, we cannot generally ensure that the overlap of the synthesized lifting operator and the device SPAM will be large or even nonzero. Instead, an alternative procedure is to synthesize SPAM-adapted operators
\begin{align}
\mathcal{A}_L &= \sum_{\ell = 1}^{\dim \phi_\lambda} |\bar{E}\rrangle\llangle\lambda^{(i)}, \ell|, \\[2 pt]
\mathcal{A}_R &= \sum_{\ell = 1}^{\dim \phi_\lambda} |\lambda^{(i')}, \ell\rrangle\llangle\bar{\rho}|,
\end{align}
where $\sket{\bar{\rho}}$ is the traceless equivalent of the native state preparation effect and similarly for $\llangle\bar{E}|$. For explicit examples of SPAM-adapted lifting operations for the computational basis, see Sections \ref{subsec:zz_numerics} and \ref{subsec:ms_numerics}.

Using the traceless components of the SPAM operators is suitable when the lifting group is a unitary 2-design (such as the Clifford group or the entire unitary group), for which the irrep $\phi_\Lambda$ is simply the orthogonal complement of the trivial representation in Hilbert-Schmidt space.  For more general $F$, we would use the components of $\rho$ and $E$ that lie within the irrep $\phi_\Lambda$.  Such bespoke synthetic operators can also be combined with synthetic SPAM targeted to the irrep $\phi_\lambda$.

As an aside, a further motivation for our lifting construction is the following.  If the native SPAM operations have zero support in certain irreps of $G$, then one needs to prepare alternative pure states to be able to extract all RB decay parameters.  Often, fiducial gates are used to enlarge the space of possible SPAM operations.  However, using elements of the benchmarking group as fiducials has no effect on the RB signal because the overlaps between SPAM and irrep projectors determine the coefficients of the exponential decays, and the projectors are invariant under conjugation by group elements.  Therefore, the fiducials must consist of gates that lie \emph{outside} the benchmarking group.

\subsection{Sample Complexity}

We now examine the sample complexity of our protocol.  The argument closely follows that for character RB with multiplicity \cite{Claes}, so it is useful to review that argument here.

\subsubsection{Character RB}

In the case of character RB, the quantity of interest is
\begin{equation}
p_m^{\lambda'} = (\dim\phi_{\lambda'})\mathbb{E}_{\vec{g}\in G^m, h\in H}[\chi_{\lambda'}(h)^\ast\sbra{E}\widetilde{\mathcal{G}}_\text{inv}\widetilde{\mathcal{G}}_m\cdots \widetilde{\mathcal{G}}_2\widetilde{\mathcal{G}_1\mathcal{H}}\sket{\rho}],
\end{equation}
which can be complex-valued if the character function is complex.  Since the magnitude of the character function $\chi_{\lambda'}(h)$ is bounded above by $\dim\phi_{\lambda'}$, estimating $p_m^{\lambda'}$ entails sampling from a random variable that is bounded in absolute value by $(\dim\phi_{\lambda'})^2$.  The standard deviation of this random variable is likewise bounded by $(\dim\phi_{\lambda'})^2$, so in the limit of large sample size $N$, the standard deviation of the sample mean is bounded by $(\dim\phi_{\lambda'})^2/\sqrt{N}$.  The sample mean itself, in the limit of zero noise (where $\mathcal{E}$ is the identity channel), is
\begin{equation}
\lim_{\mathcal{E}\to \operatorname{id}} p_m^{\lambda'} = \sbra{E}\Pi_{\lambda'}\sket{\rho}.
\end{equation}
Therefore, the uncertainty of the estimate relative to the sample mean is
\begin{equation}
\lesssim \frac{(\dim\phi_{\lambda'})^2}{\sbra{E}\Pi_{\lambda'}\sket{\rho}\sqrt{N}}.
\end{equation}
This is the standard argument from \cite{Claes}.

If the representation $\lambda'$ is (pseudo)real, then the character function $\chi_{\lambda'}(h)$ is real, so estimating $p_m^{\lambda'}$ entails sampling from a real rather than a complex random variable.  In this case, we can derive a tighter bound, following \cite{Fan:2024puh}.  We are sampling from a mixture (with respect to $\vec{g}$ and $h$) of rescaled Bernoulli random variables.  The mean and variance of a mixture of one-dimensional distributions with weights $w_i$, means $\mu_i$, and variances $\sigma_i^2$ are
\begin{equation}
\mu = \sum_i w_i\mu_i, \qquad \sigma^2 = \sum_i w_i(\sigma_i^2 + \mu_i^2) - \mu^2.
\end{equation}
If the components are rescaled Bernoulli random variables with real scale factors $\alpha_i$ and whose means and variances are therefore $\alpha_i\mu_i$ and $\alpha_i^2\mu_i(1 - \mu_i)$ for some $0\leq \mu_i\leq 1$, then we have
\begin{equation}
\mu = \sum_i w_i\alpha_i\mu_i, \qquad \sigma^2 = \sum_i w_i[\alpha_i^2\mu_i(1 - \mu_i) + \alpha_i^2\mu_i^2] - \mu^2 = \sum_i w_i\alpha_i^2\mu_i - \mu^2.
\end{equation}
While the above formulas hold for discrete mixtures, similar ones hold for continuous mixtures.  Thus our real random variable has mean
\begin{equation}
\mu = (\dim\phi_{\lambda'})\mathbb{E}_{h\in H}[\chi_{\lambda'}(h)\mathbb{E}_{\vec{g}\in G^m}[\sbra{E}\widetilde{\mathcal{G}}_\text{inv}\widetilde{\mathcal{G}}_m\cdots \widetilde{\mathcal{G}}_2\widetilde{\mathcal{G}_1\mathcal{H}}\sket{\rho}]]
\end{equation}
(no complex conjugate on $\chi_{\lambda'}(h)$) and variance
\begin{align}
\sigma^2 &= (\dim\phi_{\lambda'})^2\mathbb{E}_{h\in H}[\chi_{\lambda'}(h)^2\mathbb{E}_{\vec{g}\in G^m}[\sbra{E}\widetilde{\mathcal{G}}_\text{inv}\widetilde{\mathcal{G}}_m\cdots \widetilde{\mathcal{G}}_2\widetilde{\mathcal{G}_1\mathcal{H}}\sket{\rho}]] - \mu^2 \\
&\leq (\dim\phi_{\lambda'})^2\mathbb{E}_{h\in H}[\chi_{\lambda'}(h)^2\mathbb{E}_{\vec{g}\in G^m}[\sbra{E}\widetilde{\mathcal{G}}_\text{inv}\widetilde{\mathcal{G}}_m\cdots \widetilde{\mathcal{G}}_2\widetilde{\mathcal{G}_1\mathcal{H}}\sket{\rho}]] \\
&\leq (\dim\phi_{\lambda'})^2\mathbb{E}_{h\in H}[\chi_{\lambda'}(h)^2] \\
&= (\dim\phi_{\lambda'})^2,
\end{align}
where we have used character orthonormality in the last step.  This is a square-root improvement over the previous bound.

In summary, for character RB with multiplicity to be efficient, the following should hold \cite{Claes}:
\begin{itemize}
\item The number of inequivalent irreps in the superoperator representation of $G$ should be small.  This ensures that we need to iterate through a small number of target irreps $\lambda$ (i.e., that we need to estimate a small number of RB signals).
\item Each inequivalent irrep should occur with small multiplicity.  This ensures that each RB signal can be fit to a multi-exponential function with few parameters.
\item For each inequivalent irrep $\lambda$, it should be possible to choose a character group $H$ with a corresponding irrep $\lambda'$ such that $\dim\phi_{\lambda'}$ is small.  This ensures that estimating each RB signal requires running a reasonable number of RB experiments.
\end{itemize}
For this protocol to be scalable, the above quantities should grow slowly with the number of qubits $n$ (or, in the case of multiplicities, be bounded by a small constant).

For efficiency, one wishes to maximize the matrix element $\sbra{E}\Pi_{\lambda'}\sket{\rho}$.  Among mixed states, the optimal initial states are easily seen to be pure states \cite{Claes}, but this reasoning does not apply to synthetic states, which are unphysical (non-convex) combinations of pure states \cite{Fan:2024puh}.  One of the primary challenges of this procedure is fitting multi-exponential decays.

\subsubsection{Lifted RB}

In the case of lifted RB, the quantity of interest is
\begin{equation}
p_m^\Lambda = (\dim\phi_\Lambda)^2\mathbb{E}_{\vec{g}\in G^m, f_L\in F, f_R\in F}[(\mathcal{F}_L|\mathcal{A}_L)(\mathcal{F}_R|\mathcal{A}_R)\sbra{E}\widetilde{\mathcal{F}_L\mathcal{G}_\text{inv}}\widetilde{\mathcal{G}}_m\cdots \widetilde{\mathcal{G}}_2\widetilde{\mathcal{G}_1\mathcal{F}_R}\sket{\rho}],
\end{equation}
which is generally complex-valued.  We choose $\mathcal{A}_L = Q_{\lambda, i_L, i_0}$ and $\mathcal{A}_R = Q_{\lambda, i_0', i_R}$.  In the limit of zero noise, this becomes
\begin{equation}
\lim_{\mathcal{E}\to \operatorname{id}} p_m^\Lambda = \sbra{E}Q_{\lambda, i_L, i_R}\sket{\rho}\delta_{i_0 i_0'}.
\end{equation}
The entries of a unitary matrix are bounded in absolute value by 1.  Hence estimating $p_m^\Lambda$ entails sampling from a random variable that is bounded in absolute value by $(\dim\phi_\Lambda)^2(\dim\phi_\lambda)^2$.  In the limit of large sample size $N$, the standard deviation of the sample mean is bounded by this quantity times $1/\sqrt{N}$.  We only consider the relative uncertainty for $i_0 = i_0'$, which is
\begin{equation}
\lesssim \frac{(\dim\phi_\Lambda)^2(\dim\phi_\lambda)^2}{\sbra{E}Q_{\lambda, i_L, i_R}\sket{\rho}\sqrt{N}}.
\end{equation}
The numerator of this bound is generally much larger than that for character RB with multiplicities, but in our case, we do not have to contend with fitting multi-exponential decays.  We wish to maximize the overlap $\sbra{E}Q_{\lambda, i_L, i_R}\sket{\rho}$.  This can be done using synthetic SPAM.

For efficiency, the following should hold:
\begin{itemize}
\item The number of inequivalent irreps in the superoperator representation of $G$ should be small, for the same reason as before.
\item Each inequivalent irrep should occur with small multiplicity because the number of RB experiments that we must run for a given target irrep grows quadratically with the multiplicity.
\item For each inequivalent irrep $\lambda$, it should be possible to choose a lifting group $F$ with a corresponding irrep $\Lambda$ such that $\dim\phi_{\Lambda}$ is small.
\end{itemize}
For this protocol to be scalable, the above quantities should grow slowly with the number of qubits $n$, but we need not require that the multiplicities be bounded by a constant because our bottleneck is not fitting multi-exponential decays.  In particular, our protocol is not scalable if we simply choose $F$ to be the entire unitary group because in this case, $\dim\phi_\Lambda = 4^n - 1$ where $\Lambda$ is the adjoint representation of $SU(2^n)$.

\section{Practical Implementation}

Having discussed the framework of lifted RB, we now turn to its practical implementation.

\subsection{Theoretical Setup}

The first step is to identify a gate set, a benchmarking group $G$, and a lifting group $F$. The gate set should be able to compile both $G$ and $F$, and the gate sequence that implements any group element should be easy to compute classically. In our examples, $G$ and $F$ are both subgroups of the unitary group on $n$ qubits, $SU(2^n)$. (However, our formalism applies more generally to qudits and potentially also continuous variables.) We discuss some strategies for choosing the lifting group in Appendix \ref{app:lifting}. The goal is to measure the average fidelity of $G$, which is assumed to have multiplicity.  Given the character tables of $G$ and $F$, the relevant multiplicities can be calculated via
\begin{equation}
n_\lambda = \mathbb{E}_{g\in G}[\chi_\lambda(g)^\ast\Tr\mathcal{G}(g)],
\end{equation}
which is a consequence of the formula \eqref{projectorformula} for isotypic projectors.

Fix a target irrep $\phi_\lambda$ of $G$, which occurs with some multiplicity $n_\lambda$ in the superoperator representation $\mathcal{G}$.  We label the copies by $\lambda^{(i)}$ for $i = 1, \ldots, n_\lambda$.  Each copy corresponds to a subspace $V_{\lambda^{(i)}}$ of the Hilbert-Schmidt space $\mathcal{L}(\mathscr{H})$.  For each $i$, choose an orthonormal basis
\begin{equation}
\{|\lambda^{(i)}, \ell\rrangle \,|\, \ell = 1, \ldots, \dim\phi_\lambda\}
\end{equation}
of operators (superkets) for $V_{\lambda^{(i)}}$ such that
\begin{equation}
\llangle\lambda^{(i)}, \ell|\mathcal{G}(g)|\lambda^{(i)}, \ell'\rrangle = \llangle\lambda^{(i')}, \ell|\mathcal{G}(g)|\lambda^{(i')}, \ell'\rrangle
\end{equation}
for all elements $g\in G$, all $i, i' = 1, \ldots, n_\lambda$, and all $\ell, \ell' = 1, \ldots, \dim\phi_\lambda$.  In other words: choose orthonormal bases for the $V_{\lambda^{(i)}}$ such that the superoperator matrix elements of $\mathcal{G}(g)$ in these bases are identical for all $V_{\lambda^{(i)}}$.  It is always possible to choose such bases for the $V_{\lambda^{(i)}}$ because the representations $\lambda^{(i)}$ are isomorphic.  Finally, for each pair $(i, i')$, define the operator
\begin{equation}
Q_{\lambda, i, i'}\equiv \sum_{\ell=1}^{\dim\phi_\lambda} |\lambda^{(i)}, \ell\rrangle\llangle\lambda^{(i')}, \ell|
\end{equation}
in terms of outer products of the associated basis elements.  This is an explicit isomorphism from $V_{\lambda^{(i')}}$ to $V_{\lambda^{(i)}}$.  We will need the operators $Q_{\lambda, i, i'}$ for the data analysis procedure.

How should we construct the basis vectors $|\lambda^{(i)}, \ell\rrangle$?  They are all eigenvectors of $\Pi_\lambda$ with eigenvalue 1, but when $\dim\phi_\lambda > 1$, they must satisfy additional constraints.  Here is a recipe.  First, construct the isotypic projector $\Pi_\lambda$ using the standard character-based formula \eqref{projectorformula}.  The image of $\Pi_\lambda$ is the isotypic component $V_\lambda$ of the irrep $\phi_\lambda$:
\begin{equation}
V_\lambda\cong V_{\lambda^{(1)}}\oplus \cdots\oplus V_{\lambda^{(n_\lambda)}}\cong \mathbb{C}^{n_\lambda}\otimes \widetilde{V}_\lambda,
\end{equation}
where $\widetilde{V}_\lambda$ is an abstract carrier space for $\phi_\lambda$.  Let $\mathcal{G}(g)|_\lambda$ denote the restriction of the superoperator representation of $g$ to $V_\lambda$.  Consider the intertwiner space
\begin{equation}
\operatorname{Hom}_G(\widetilde{V}_\lambda, V_\lambda) = \{T : \widetilde{V}_\lambda\to V_\lambda \text{ such that } \mathcal{G}(g)|_\lambda T = T\phi_\lambda(g) \text{ for all } g\in G\},
\end{equation}
which, by Schur's lemma, has dimension $n_\lambda$.  Choose an orthonormal basis $T_1, \ldots, T_{n_\lambda}$ for this space with respect to the normalized Hilbert-Schmidt inner product:
\begin{equation}
\frac{\tr(T_i^\dag T_j)}{\dim\phi_\lambda} = \delta_{ij}.
\end{equation}
Schur's lemma then implies that
\begin{equation}
T_i^\dag T_j = \delta_{ij}I_{\widetilde{V}_\lambda}.
\end{equation}
Finally, choose an orthonormal basis $\{|\ell\rrangle \,|\, \ell = 1, \ldots, \dim\phi_\lambda\}$ for $\widetilde{V}_\lambda$ and let
\begin{equation}
|\lambda^{(i)}, \ell\rrangle = T_i|\ell\rrangle.
\end{equation}
In this basis for $V_\lambda$, we have $\mathcal{G}(g)|_\lambda = \phi_\lambda(g)\oplus \cdots\oplus \phi_\lambda(g)$.  When $n_\lambda > 1$, our non-canonical choice of basis for $\operatorname{Hom}_G(\widetilde{V}_\lambda, V_\lambda)$ is equivalent to a choice of basis for the multiplicity space $\mathbb{C}^{n_\lambda}$.

\subsection{Experimental Protocol}

We assume that we have access to a native state preparation $|0\rangle^{\otimes n}$ and a native projective measurement that spans all binary outcomes for each qubit (in other words, we assume computational 0 and measurement in the computational basis).  We also assume for simplicity that the lifting group is a unitary 2-design.

The experimental and analysis procedure, in algorithmic form, is as follows.  Fix two indices $i_L, i_R\in \{1, \ldots, n_\lambda\}$.  For each desired sequence length $m\in M$ (where $M$ is the set of sequence lengths), do the following:
\begin{itemize}
\item Choose a sample size $N$ (this is the number of shots).  To get equally good statistics for each $m$, we choose $N$ to increase with $m$.\footnote{More precisely: for the estimated RB survival probability at each depth $m$ to have approximately the same standard error, the central limit theorem says that we should take $N$ to scale linearly with the sequence-to-sequence variance of the measured RB outcome at depth $m$.  For standard Clifford RB with generic Markovian and gate-independent noise, a good rule of thumb is $N\sim m^2$ \cite{RB-confidence}.}
\item Define $n_\lambda^2$ complex variables $x_{ii'}$ (where $i, i' = 1, \ldots, n_\lambda$) and initialize them all to 0.
\item For each $k = 1, \ldots, N$:
\begin{itemize}
\item Choose random group elements $f_\rightarrow, f_\leftarrow\in F$ and $g_1, \ldots, g_m\in G$ and apply the gates $g_1 f_\rightarrow$, $g_2$, \ldots, $g_m$, $f_\leftarrow(g_m\cdots g_1)^{-1}$ (which are compiled into native gates) to the $|0\rangle$ state, in that order.
\item Measure the resulting state in the computational basis.  If the result is not 0, do nothing.  If the result is 0, then for all $i, i' = 1, \ldots, n_\lambda$, increment $x_{ii'}$ by the quantity
\begin{equation}
(f_\leftarrow|Q_{\lambda, i_L, i})(f_\rightarrow|Q_{\lambda, i', i_R}),
\end{equation}
where $(f|Q)\equiv \Tr(f^\dag Q)$ denotes the Hilbert-Schmidt inner product between the superoperator $f$ and the superoperator $Q$ ($\Tr$ is the superoperator trace).\footnote{In this step, the $Q$'s can be replaced by SPAM-adapted lifting operators.}
\end{itemize}
\item Rescale all of the $n_\lambda^2$ complex variables $x_{ii'}$ by $(4^n - 1)^2/N$, where $4^n - 1$ is the dimension of the adjoint representation of $SU(2^n)$.
\item Let $X^{(m)}$ be the $n_\lambda\times n_\lambda$ matrix with entries $x_{ii'}$.  Store this matrix.
\end{itemize}
Finally, fit the collection of matrices $\{X^{(m)} \,|\, m\in M\}$ to a matrix exponential function of the form $m\mapsto cM_\lambda^m$ by minimizing a sum of matrix norms:
\begin{equation}
\widehat{M}_\lambda = \underset{A}{\operatorname{argmin}} \min_c \sum_m \big| c A^m - X^{(m)} \big|.
\end{equation}
The eigenvalues of $\widehat{M}_\lambda$ are our estimates for the decay parameters of the irrep $\phi_\lambda$.

We can simplify the experimental procedure as follows.  An experimentalist, having chosen random group elements $f_\rightarrow, g_1, \ldots, g_m, f_\leftarrow$, can simply apply the gates $f_\rightarrow$, $g_2$, \ldots, $g_m$, $f_\leftarrow$ to the $|0\rangle$ state.  Having done so, one would increment $x_{ii'}$ by the quantity
\begin{equation}
(f_\leftarrow g_m\cdots g_1|Q_{\lambda, i_L, i})(g_1^{-1}f_{\rightarrow}|Q_{\lambda, i', i_R}).
\end{equation}
This eliminates the need for compilation and even inversion of physical gates.  Note that the gate $g_1$ is never physically applied; it appears entirely in classical post-processing, and it can be averaged over separately from the other gates.  This perspective (which aligns with that of filtered RB) transfers the difficulty in the experiment design to the data analysis, making the procedure easier to implement experimentally.

\subsection{Gate Fidelities and Error Rates}

Using the union of the RB decay parameters over all $\lambda$, we can estimate the average process fidelity (equivalently, entanglement fidelity) of the benchmarking group via
\begin{equation}
\widehat{F}_\text{process} = d^{-2}\sum_\lambda (\dim\phi_\lambda)\tr\widehat{M}_\lambda.
\end{equation}
Indeed, computing the average gate fidelity requires knowing all of the separate decay rates in the RB signal.\footnote{See, however, earlier work on approximate RB for arbitrary finite groups \cite{RB-approximate-1, RB-approximate-2, RB-approximate-3}, which attempts to make do with less.}  For a multiplicity-free benchmarking group $G$, we can write a $G$-twirled error channel $\mathcal{E}$ in the form
\begin{equation}
\mathbb{E}_{g\in G}[\mathcal{G}^{-1}\mathcal{E}\mathcal{G}] = \sum_\lambda \gamma_\lambda(\mathcal{E})\Pi_\lambda, \qquad \gamma_\lambda(\mathcal{E}) = \frac{\Tr(\Pi_\lambda\mathcal{E})}{\Tr(\Pi_\lambda)}.
\end{equation}
The entanglement fidelity and average gate fidelity of $G$ are given by \cite{Nielsen, Hashim}
\begin{equation}
F_e = d^{-2}\sum_\lambda \Tr(\Pi_\lambda)\gamma_\lambda(\mathcal{E}), \qquad \overline{F} = \frac{dF_e + 1}{d + 1}.
\end{equation}
In the presence of multiplicity, the formula for entanglement fidelity generalizes to \cite{Claes}
\begin{equation}
F_e = d^{-2}\sum_\lambda (\dim\phi_\lambda)\sum_{i=1}^{n_\lambda} \gamma_{\lambda, i}(\mathcal{E}).
\end{equation}
Since $F_e\leq 1$, it follows that $F_e\leq \overline{F}\leq 1$.

We also record some useful formulas for the error rates within irrep sectors.  As usual, let $\lambda$ label the superoperator irreps of $G$.  If $G$ is multiplicity-free, then the normalized error probability (Choi weight) of an arbitrary error channel $\mathcal{E}$ in irrep $\lambda$ is given by
\begin{equation}
p_\lambda = \frac{\Tr(\Pi_\lambda J(\mathcal{E}))}{d},
\label{choiweight}
\end{equation}
where $J(\mathcal{E})$ is the corresponding Choi matrix and $d$ is the dimension of the Hilbert space.  Letting $\lambda = 0$ label the trivial irrep spanned by the identity operator, we have $p_0 = F_e$.  In terms of Kraus operators, we have
\begin{equation}
\mathcal{E}(\rho) = \sum_a K_a\rho K_a^\dag, \qquad J(\mathcal{E}) = \sum_a |K_a\rrangle\llangle K_a|,
\end{equation}
and therefore
\begin{equation}
p_\lambda = \frac{1}{d}\sum_a \llangle K_a|\Pi_\lambda|K_a\rrangle,
\end{equation}
which is independent of the choice of Kraus representation.  Note that $\llangle\mathcal{O}|\Pi_\lambda|\mathcal{O}\rrangle$ is the squared norm of the component of the operator $\mathcal{O}$ within irrep $\lambda$.  In the case of a unitary channel $\rho\mapsto U\rho U^\dag$, we have simply
\begin{equation}
p_\lambda = \frac{\llangle U|\Pi_\lambda|U\rrangle}{d}.
\end{equation}
In particular, for \emph{small} coherent errors where $U = e^{-iH} = \mathds{1} - iH + O(H^2)$ and
\begin{equation}
\rho\mapsto \rho - i[H, \rho] + \left(H\rho H - \frac{1}{2}\{H^2, \rho\}\right) + O(H^3),
\end{equation}
we have to leading order in $H$ that
\begin{equation}
p_{\lambda > 0} = \frac{\llangle H|\Pi_\lambda|H\rrangle}{d} + O(H^3),
\end{equation}
where we have used $\Pi_{\lambda > 0}|\mathds{1}\rrangle = 0$.  In the case of multiplicity, using the appropriate formula for the $G$-twirl of $J(\mathcal{E})$ shows that the scalar $p_\lambda$ is replaced by a positive-semidefinite matrix $P_\lambda$ on the multiplicity space of $\phi_\lambda$:
\begin{equation}
P_\lambda = \frac{\dim\phi_\lambda}{d}M_\lambda(J(\mathcal{E})),
\end{equation}
where $M_\lambda(J(\mathcal{E}))$ is the $n_\lambda\times n_\lambda$ matrix with entries
\begin{equation}
[M_\lambda(J(\mathcal{E}))]_{ii'} = \frac{\Tr(Q_{\lambda, i', i}J(\mathcal{E}))}{\dim\phi_\lambda}.
\end{equation}
The total probability associated with the entire isotypic component of the irrep $\phi_\lambda$ is then
\begin{equation}
p_\lambda = \tr P_\lambda = \frac{\Tr(\Pi_\lambda J(\mathcal{E}))}{d},
\end{equation}
as in \eqref{choiweight}, where $\Pi_\lambda$ is now understood to be an isotypic projector.

\section{Applications}

Our ``lifted'' version of synthetic RB opens the door to many applications, including improved subspace and leakage RB in trapped-ion quantum computers \cite{Sutherland:2025shv}.  Subspace RB \cite{Baldwin} seeks to characterize gates that preserve a subspace of the Hilbert space, while leakage RB \cite{Wood} characterizes leakage errors into non-computational subspaces of the Hilbert space.  Gates that preserve a proper subspace cannot form a unitary 2-design (or even a 1-design), and their superoperator representations are never multiplicity-free because the projector onto the common invariant subspace yields a copy of the trivial representation distinct from that spanned by the identity operator \cite{Claes}.

Note that the mathematical structure of \emph{symmetric} subspace RB as formulated in \cite{Baldwin, Claes}, which applies to one- and two-qubit gates that are invariant under SWAP, is too simple to exhibit the generality of our techniques.  There, only the trivial irrep occurs with multiplicity, and since it has multiplicity two, extracting the relevant signal only requires fitting a single exponential decay with a constant offset.  This is not a true ``multi-exponential'' problem and does not require sophisticated techniques.

We examine how to implement the lifting group $F$ in some concrete two-qubit scenarios.  Our primary interest lies in benchmarking certain two-qubit M{\o}lmer-S{\o}rensen (MS) gates of the form \cite{Sutherland:2025shv}
\begin{equation}
\operatorname{MS}(\theta, \phi) = \left[\begin{array}{cccc}
\cos(\theta/2) & 0 & 0 & ie^{-2i\phi}\sin(\theta/2) \\
0 & \cos(\theta/2) & i\sin(\theta/2) & 0 \\
0 & i\sin(\theta/2) & \cos(\theta/2) & 0 \\
ie^{2i\phi}\sin(\theta/2) & 0 & 0 & \cos(\theta/2)
\end{array}\right].
\label{msgates}
\end{equation}
Any two such gates commute within the odd-parity subspace.  We assume that we are able to prepare and measure computational basis states ($|00\rangle$, $|01\rangle$, $|10\rangle$, $|11\rangle$).

Note that the gates \eqref{msgates} are examples of matchgates.  The two-qubit matchgate group is simply the subgroup of $U(2)_\text{even}\times U(2)_\text{odd}$ where the even and odd blocks, which act within the even- and odd-parity subspaces, have equal determinant.  Projectively, it is
\begin{equation}
\frac{SU(2)_\text{even}\times SU(2)_\text{odd}}{\{(I, I), (-I, -I)\}}\cong SO(4).
\end{equation}
The matchgate group on $n$ qubits, arranged in a line, is generated by nearest-neighbor two-qubit matchgates and projectively isomorphic to $SO(2n)$.  Like the Clifford group, the matchgate group is efficiently classically simulable and becomes universal with the addition of one extra gate.  Also like the Clifford group, the matchgate group can be efficiently benchmarked \cite{Claes, Helsen2022matchgate}.  In the case of the matchgate group, this is because the irreps of its conjugation representation occur with multiplicity at most two.  Here, we do not consider the full matchgate group, but only subgroups thereof, which present greater challenges due to their potentially larger multiplicities.

Let us define the single-qubit rotations $R_p(\theta) = e^{-i(\theta/2)P}$ with $p \linebreak[1] \in \linebreak[1] \{x, y, z\}$.  In particular, we write
\begin{equation}
R_x = R_x(-\pi/2) = \frac{1}{\sqrt{2}}\left[\begin{array}{cc}
1 & i \\
i & 1
\end{array}\right], \qquad
R_y = R_y(-\pi/2) = \frac{1}{\sqrt{2}}\left[\begin{array}{cc}
1 & 1 \\
-1 & 1
\end{array}\right].
\end{equation}
Note that
\begin{equation}
R_x^4 = R_y^4 = (R_x^2 R_y^2)^2 = -I.
\end{equation}
Consider the two-qubit unitary gates
\begin{equation}
M_{ab}\equiv R_a^{00, 11}\oplus R_b^{01, 10}
\end{equation}
where $a, b\in \{x, y\}$, which we have written in terms of single-qubit rotations within the even- and odd-parity subspaces.  Concretely, let us write
\begin{align}
M_x &\equiv M_{xx} = R_x^{00, 11}\oplus R_x^{01, 10} = \frac{1}{\sqrt{2}}\left[\begin{array}{cccc}
1 & 0 & 0 & i \\
0 & 1 & i & 0 \\
0 & i & 1 & 0 \\
i & 0 & 0 & 1
\end{array}\right], \\
M_y &\equiv M_{yx} = R_y^{00, 11}\oplus R_x^{01, 10} = \frac{1}{\sqrt{2}}\left[\begin{array}{cccc}
1 & 0 & 0 & 1 \\
0 & 1 & i & 0 \\
0 & i & 1 & 0 \\
-1 & 0 & 0 & 1
\end{array}\right].
\end{align}
We also have
\begin{align}
M_{yy} &= R_y^{00, 11}\oplus R_y^{01, 10} = \frac{1}{\sqrt{2}}\left[\begin{array}{cccc}
1 & 0 & 0 & 1 \\
0 & 1 & 1 & 0 \\
0 & -1 & 1 & 0 \\
-1 & 0 & 0 & 1
\end{array}\right], \\
M_{xy} &= R_x^{00, 11}\oplus R_y^{01, 10} = \frac{1}{\sqrt{2}}\left[\begin{array}{cccc}
1 & 0 & 0 & i \\
0 & 1 & 1 & 0 \\
0 & -1 & 1 & 0 \\
i & 0 & 0 & 1
\end{array}\right].
\end{align}
Let us also write
\begin{equation}
M_{zz} = e^{i(\pi/4)Z\otimes Z} =
\left[\begin{array}{cccc}
e^{i\pi/4} & 0 & 0 & 0 \\
0 & e^{-i\pi/4} & 0 & 0 \\
0 & 0 & e^{-i\pi/4} & 0 \\
0 & 0 & 0 & e^{i\pi/4}
\end{array}\right],
\end{equation}
which should \emph{not} be confused with the gate $R_z^{00, 11}\oplus R_z^{01, 10}$ where
\begin{equation}
R_z = R_z(-\pi/2) = \left[\begin{array}{cc}
e^{i\pi/4} & 0 \\
0 & e^{-i\pi/4}
\end{array}\right].
\end{equation}
Note that the following gates are two-qubit Clifford gates:
\begin{equation}
M_x = M_{xx} = e^{i(\pi/4)X\otimes X}, \qquad M_{yy} = e^{i(\pi/4)Y\otimes X}, \qquad M_{zz} = e^{i(\pi/4)Z\otimes Z}.
\end{equation}
However, the following gates are not:
\begin{equation}
M_y = M_{yx}, \qquad M_{xy}.
\end{equation}
The Pauli transfer matrix (PTM) representation of any Clifford gate is a signed permutation matrix.  Although $M_y$ is not a Clifford gate with respect to the original two-qubit Pauli group, it acts as a one-qubit Clifford gate in each parity sector.

In our examples, we study the quality of fidelity estimation using lifted randomized benchmarking. We specifically analyze the implementation of a ZZ gate and a pair of in-phase and quadrature MS gates.\footnote{Randomized benchmarking is only sensitive to the adjoint image of the benchmarking group, i.e., the group generated by the corresponding unitary channels.  For example, the $n$-qubit Pauli group is non-abelian, but its projective version is the abelian group $\mathbb{Z}_2^{2n}$.} We find that the procedure can reliably estimate the fidelity of these two-qubit gates with reasonable shot overhead.

\subsection{Example: ZZ Gate} \label{subsec:zz}

In this example, the benchmarking group is the cyclic group $\mathbb{Z}_8$ generated by the single ZZ gate $M_{zz}$, which, at the level of superoperators, is represented as $\mathbb{Z}_4$.  Since the benchmarking group is abelian, it only has 1D irreps.

The irreps that occur in the superoperator representation are as follows (below, we leave tensor products implicit):
\begin{itemize}
\item The trivial representation $\phi_0$ has dimension 1 and multiplicity 8.  Its isotypic component has basis
\begin{equation}
\frac{1}{2}\left\{II, IZ, ZI, ZZ, XX, XY, YX, YY\right\}.
\end{equation}
\item The representation $\phi_+$, in which $M_{zz}$ acts as multiplication by $+i$, has dimension 1 and multiplicity 4.  Its isotypic component has basis
\begin{equation}
\frac{1}{2\sqrt{2}}\left\{IX + iZY, XI + iYZ, IY - iZX, YI - iXZ\right\}.
\end{equation}
\item The representation $\phi_-$, in which $M_{zz}$ acts as multiplication by $-i$, has dimension 1 and multiplicity 4.  Its isotypic component has basis
\begin{equation}
\frac{1}{2\sqrt{2}}\left\{IX - iZY, XI - iYZ, IY + iZX, YI + iXZ\right\}.
\end{equation}
\end{itemize}

Since $M_{zz}$ is a Clifford gate, an ideal lifting group for this example is the two-qubit Clifford group, which is both easy to sample from and analytically tractable.

\subsubsection{Numerical Implementation} \label{subsec:zz_numerics}

As a validation step, we implement our procedure for the ZZ gate numerically and analyze the results. We explore varying ranges of infidelity as well as circuit count, shot count, and circuit depth choices. We find that the procedure can robustly estimate the infidelity of a gate and, additionally, resolve the error channel's eigenvalues to high precision. 

In all our examples below, we consider state preparation of computational 0 and measurement in the computational basis. Accordingly, we choose the SPAM-adapted basis for our lifting operators: see Section \ref{subsec:spam_lifting}. For $\sket{00} = (\sket{II} + \sket{IZ} + \sket{ZI} + \sket{ZZ})/2$, the traceless equivalent is $(\sket{IZ} + \sket{ZI} + \sket{ZZ})/\sqrt{3}$. Therefore, we synthesize right lifting operators of the form 
\begin{equation}
\mathcal{A}_R = \sum_\ell \sket{\lambda^{(i)}, \ell}\left( \frac{\sbra{IZ} + \sbra{ZI} + \sbra{ZZ}}{\sqrt{3}} \right), 
\end{equation}
and similarly for the left lifting operator. 

\begin{figure*}
\centering
\includegraphics[width=\linewidth]{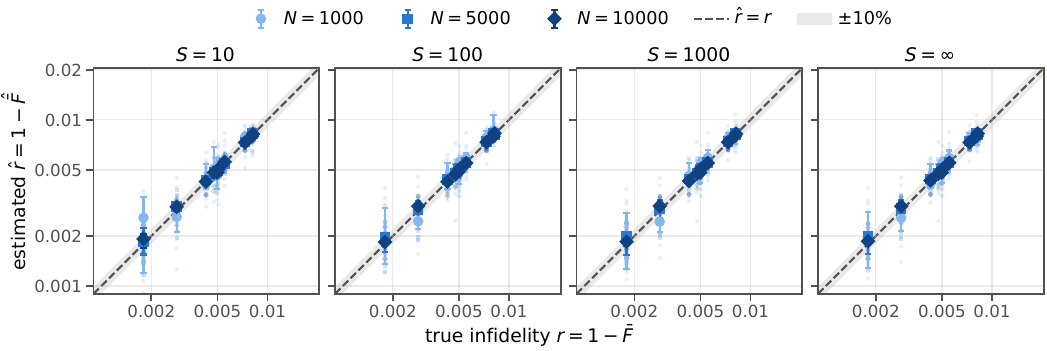}
\caption{Performance of lifted randomized benchmarking with gate-independent noise over varying noise channel infidelities, circuits per depth, and shots per circuit. These fits were constructed over the geometric interval $2^k$ with maximum depth 64, $N$ circuits per depth, and $S$ shots per circuit. Each point represents the median of 20 trials of the lifting benchmarking protocol for a given noise channel, and the error bars represent the central $68\%$ intervals. Each element of the lifting group and benchmarking group has the same noise channel. We find that for $N=1000$, some of the fits fail and produce estimates with infidelity less than 0, and for the others with greater circuits per depth, none of the fits produce negative estimates.}
\label{fig:ZZ_indepdent_fidel}
\end{figure*}

\begin{figure}
\centering
\includegraphics[width=0.5\linewidth]{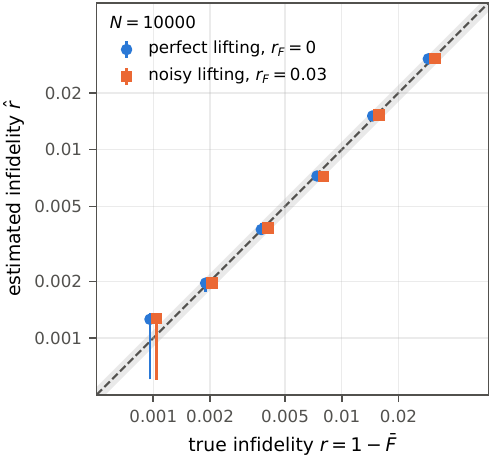}
\caption{Estimated fidelity with noisy lifting vs.\ perfect lifting. We apply our procedure for the ZZ gate with 10,000 circuits per depth and 100 shots per circuit. Each point represents the median of 20 trials of the lifting benchmarking protocol for a given noise channel, and the error bars represent the central $68\%$ intervals. In the case of perfect lifting, perfect lifting operators are applied (which means that only the last lifting operator that was not compiled into the benchmarking sequence is perfect). In the case of noisy lifting, both the first compiled and the last lifting operators contain a separate noise channel of infidelity equal to 0.03, which is larger than the benchmarking infidelity. We see that our procedure is robust to imperfections in the lifting operators.}
\label{fig:ZZ_noisy_lifting}
\end{figure}

\begin{figure*}
\centering
\includegraphics[width=\linewidth]{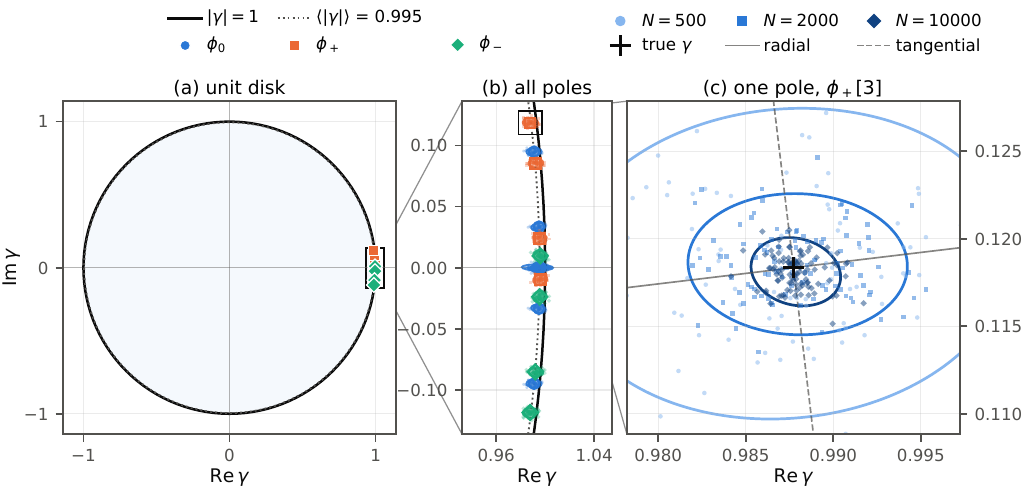}
\caption{Estimated poles of the ZZ gate error channel in the complex plane. A single noise channel was fixed at the beginning of the experiment, and each point represents an experimental implementation with that fixed noise channel. The ZZ gate benchmarking group projects a general CPTP noise process onto the irrep decomposition, and our procedure robustly measures the eigenvalues of the isotypic components. These fits were constructed over the geometric interval $2^k$ with maximum depth 64, $N$ circuits per depth, and 1000 shots per circuit.}
\label{fig:ZZ_poles}
\end{figure*}

We begin with the case of gate-independent noise, with uniform error channels decorating all benchmarking and lifting group elements. We sample a random noise channel with a given target infidelity. The noise has both incoherent Pauli stochastic components and coherent components, with about 50\% total contribution to the error budget for each. We use a geometric sequence of circuit depths in powers of 2 up to 64. We vary over the number of shots per circuit, the number of circuits per depth, and the maximum depth. We also include an ``infinite'' shot per circuit trail, which means that instead of sampling from a calculated expectation value, we just use the expectation value itself. We collect many random experiments over ranges of infidelity and repetitions in Figure \ref{fig:ZZ_indepdent_fidel} for gate independent noise, and we compare noisy and perfect lifting in Figure \ref{fig:ZZ_noisy_lifting}. Finally, we also include a plot of the spread of the eigenvalue estimates in the complex plane in Figure \ref{fig:ZZ_poles}.

\subsection{Example: MS Gates}

We now proceed to a slightly more complicated example in which the benchmarking group is non-abelian and has 2D irreps.  This is the group used for subspace leakage error randomized benchmarking (SLERB) \cite{Sutherland:2025shv}.

We consider the group generated by the matrices $M_x$ and $M_y$.  This group has $2\times 96 = 192$ elements.  The projective benchmarking group has 96 elements and 16 conjugacy classes.  It has 24 elements when restricted to the even-parity subspace.  It is not a direct product since the character table of a direct product of groups is the Kronecker product of the individual character tables.  The irrep structure and character table of the projective benchmarking group are given in \cite{Sutherland:2025shv}.  Since $M_y$ is not a Clifford gate, the SLERB group is not a subgroup of the two-qubit Clifford group.  Note that $M_x = \operatorname{MS}(\pi/2, 0)$ and $M_y = \operatorname{MS}(\pi/2, \pi/4)$.

Let us define the Bell states
\begin{equation}
|\Psi^\pm\rangle = \frac{1}{\sqrt{2}}(|01\rangle\pm |10\rangle).
\end{equation}
The six distinct irreps that occur in the superoperator representation are as follows, where we choose operator bases that are orthonormal with respect to the Hilbert-Schmidt inner product:
\begin{itemize}
\item The trivial representation $\phi_0$ has dimension 1 and multiplicity 3.  Its isotypic component has basis
\begin{equation}
\frac{1}{\sqrt{2}}(|00\rangle\langle 00| + |11\rangle\langle 11|), \qquad |\Psi^+\rangle\langle\Psi^+|, \qquad |\Psi^-\rangle\langle\Psi^-|.
\end{equation}
Alternatively, in terms of the Pauli basis, we can use
\begin{equation}
\frac{II}{2}, \qquad \frac{ZZ}{2}, \qquad \frac{XX + YY}{2\sqrt{2}}.
\end{equation}
\item The real representation $\phi_c$ has dimension 3 and multiplicity 1.  It has basis
\begin{equation}
|00\rangle\langle 11|, \qquad |11\rangle\langle 00|, \qquad \frac{1}{\sqrt{2}}(|00\rangle\langle 00| - |11\rangle\langle 11|).
\end{equation}
Alternatively, in terms of the Pauli basis, we can use
\begin{equation}
\frac{XX - YY}{2\sqrt{2}}, \qquad \frac{XY + YX}{2\sqrt{2}}, \qquad \frac{ZI + IZ}{2\sqrt{2}}.
\end{equation}
This is the irrep associated with the Clifford action of the benchmarking group on the even-parity subspace spanned by $|00\rangle, |11\rangle$.
\item The complex representation $\phi_{a, 1}$ has dimension 1 and multiplicity 1.  It has basis
\begin{equation}
|\Psi^-\rangle\langle\Psi^+|.
\end{equation}
In this representation, both $M_x$ and $M_y$ correspond to multiplication by $-i$.
\item The complex representation $\phi_{a, 2}$ has dimension 1 and multiplicity 1.  It has basis
\begin{equation}
|\Psi^+\rangle\langle\Psi^-|.
\end{equation}
In this representation, both $M_x$ and $M_y$ correspond to multiplication by $+i$.
\item The complex representation $\phi_{b, 1}$ has dimension 2 and multiplicity 2.  Its isotypic component has basis
\begin{equation}
|\Psi^+\rangle\langle 00|, \qquad |\Psi^+\rangle\langle 11|, \qquad |00\rangle\langle\Psi^-|, \qquad |11\rangle\langle\Psi^-|.
\end{equation}
Alternatively:
\begin{equation}
\frac{IX - iYZ}{2\sqrt{2}}, \qquad \frac{IY + iXZ}{2\sqrt{2}}, \qquad \frac{XI - iZY}{2\sqrt{2}}, \qquad \frac{YI + iZX}{2\sqrt{2}}.
\end{equation}
\item The complex representation $\phi_{b, 2}$ has dimension 2 and multiplicity 2.  Its isotypic component has basis
\begin{equation}
|\Psi^-\rangle\langle 00|, \qquad |\Psi^-\rangle\langle 11|, \qquad |00\rangle\langle\Psi^+|, \qquad |11\rangle\langle\Psi^+|.
\end{equation}
Alternatively:
\begin{equation}
\frac{IX + iYZ}{2\sqrt{2}}, \qquad \frac{IY - iXZ}{2\sqrt{2}}, \qquad \frac{XI + iZY}{2\sqrt{2}}, \qquad \frac{YI - iZX}{2\sqrt{2}}.
\end{equation}
\end{itemize}
The representations $\phi_{a, 1}, \phi_{a, 2}$ and $\phi_{b, 1}, \phi_{b, 2}$ form complex-conjugate pairs.  The corresponding projectors can be constructed either using the above basis elements or using the character formula \eqref{projectorformula}.  In the given bases for $\phi_{b, 1}$ and $\phi_{b, 2}$, group element superoperators (restricted to the support of either the $\phi_{b, 1}$ or the $\phi_{b, 2}$ representation) take the form
\begin{equation}
\left[\begin{array}{cccc}
a & b & 0 & 0 \\
c & d & 0 & 0 \\
0 & 0 & d & -c \\
0 & 0 & -b & a
\end{array}\right]
=
\left[\begin{array}{cccc}
1 & 0 & 0 & 0 \\
0 & 1 & 0 & 0 \\
0 & 0 & 0 & \mp 1 \\
0 & 0 & \pm 1 & 0
\end{array}\right]
\left[\begin{array}{cccc}
a & b & 0 & 0 \\
c & d & 0 & 0 \\
0 & 0 & a & b \\
0 & 0 & c & d
\end{array}\right]
\left[\begin{array}{cccc}
1 & 0 & 0 & 0 \\
0 & 1 & 0 & 0 \\
0 & 0 & 0 & \pm 1 \\
0 & 0 & \mp 1 & 0
\end{array}\right].
\end{equation}
Therefore, if we choose the ordered bases
\begin{equation}
\rho_1^{(1)} = |\Psi^+\rangle\langle 00|, \qquad \rho_2^{(1)} = |\Psi^+\rangle\langle 11|, \qquad \rho_3^{(1)} = |11\rangle\langle\Psi^-|, \qquad \rho_4^{(1)} = -|00\rangle\langle\Psi^-|
\label{orderedbasis1}
\end{equation}
for $\phi_{b, 1}$ and
\begin{equation}
\rho_1^{(2)} = |\Psi^-\rangle\langle 00|, \qquad \rho_2^{(2)} = |\Psi^-\rangle\langle 11|, \qquad \rho_3^{(2)} = |11\rangle\langle\Psi^+|, \qquad \rho_4^{(2)} = -|00\rangle\langle\Psi^+|
\label{orderedbasis2}
\end{equation}
for $\phi_{b, 2}$, then group element superoperators take the form of block diagonal $4\times 4$ matrices with identical $2\times 2$ unitary blocks.  In other words, group element superoperators take the form $I\otimes U$ for some unitary $U$.  This form is preserved under conjugation by $V_1\otimes V_2$ for unitaries $V_1$ and $V_2$.  The matrix elements are independent of the choice of $V_1$, i.e., of the choice of basis for the multiplicity space.  We have thus identified a suitable basis for the multiplicity analysis, which is a refinement of the basis used in \cite{Sutherland:2025shv}.

In this example, our techniques can be used to extract the individual quality parameters for (1) the multiplicity-3 irrep $\phi_0$ of dimension 1, (2) the multiplicity-2 irrep $\phi_{b, 1}$ of dimension 2, and (3) the multiplicity-2 irrep $\phi_{b, 2}$ of dimension 2.  The other irreps do not occur with multiplicity and can be handled with ordinary character RB.

We use $SU(4)$ as the lifting group (for a detailed justification, see Appendix \ref{app:msgates}).  Since this is a two-qubit example, scalability is not an issue.

\subsubsection{Numerical Implementation} \label{subsec:ms_numerics}

\begin{figure}
\centering
\includegraphics[width=\linewidth]{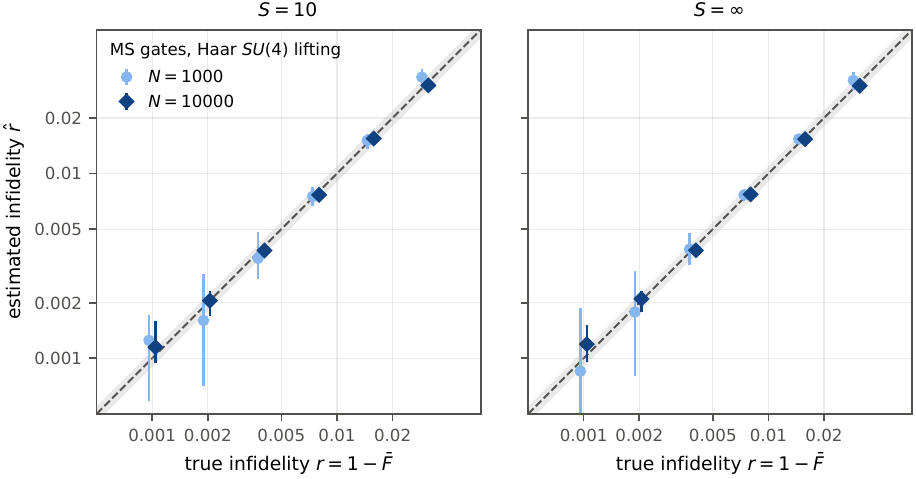}
\caption{MS gate benchmarking results for gate-independent noise. Each point represents the median of 20 trials of the lifting benchmarking protocol over a given noise channel, and the error bars represent the central $68\%$ intervals. The error channels consist of equal parts Pauli stochastic and coherent noise.}
\label{fig:MS_parity}
\end{figure}

\begin{figure}
\centering
\includegraphics[width=\linewidth]{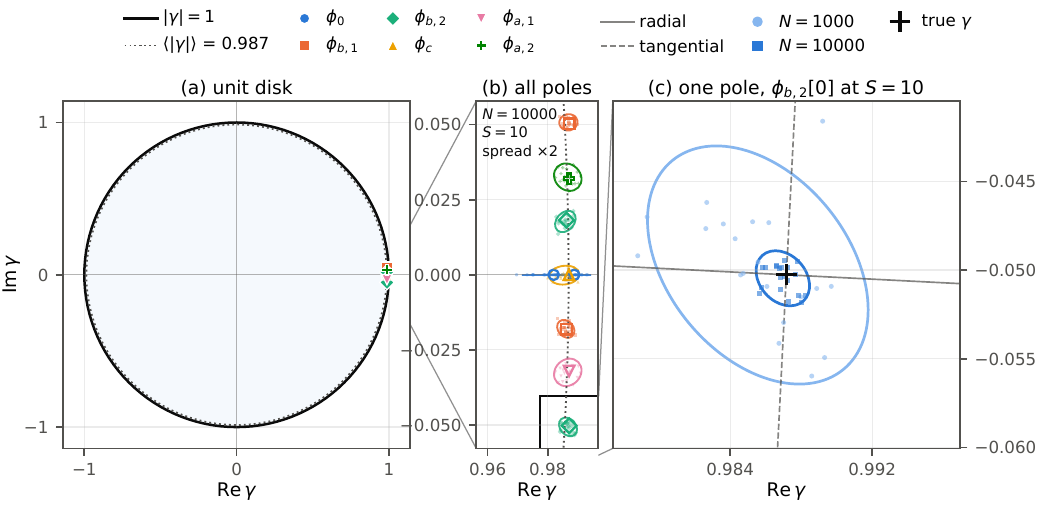}
\caption{Estimated poles of the MS gate error channel in the complex plane. A single noise channel was fixed at the beginning of the experiment, and each point represents an experimental implementation with that fixed noise channel. The MS gate benchmarking group projects a general CPTP noise process onto the irrep decomposition, and our procedure robustly measures the eigenvalues of the isotypic components. These fits were constructed over the geometric interval $2^k$ with maximum depth 64, $N$ circuits per depth, and 1000 shots per circuit.}
\label{fig:MS_poles}
\end{figure}

We implemented the MS gate protocol numerically to validate our analysis. As with the ZZ numerical experiments, we consider native state preparation and measurement in the computational basis, and we use SPAM-adapted lifting operators. Our results are shown in Figures \ref{fig:MS_parity} and \ref{fig:MS_poles}. These results indicate that our protocol can be deployed to benchmark non-abelian groups with multiplicity. 

\section{Discussion}

Accurately measuring the quality of a set of gates is a fundamental problem in the engineering of high-quality quantum processors. One of the most well-studied approaches to this problem is randomized benchmarking, which composes sequences of operations that represent group actions on density matrices. When the representation of the benchmarking group has multiplicity, each symmetry sector can contribute multiple exponential decays to the experimentally measured success probability. Our work provides a complete solution to the problem of benchmarking non-multiplicity-free groups by lifting the group structure to allow for isolation of single-exponential decays. This solution is intrinsic in the sense that it exploits the symmetry structure of a group of quantum operations rather than merely classical techniques.

The prior state of the art for benchmarking groups with multiplicity \cite{Claes} involves extracting individual eigenvalue decays from matrix exponential decays using sophisticated signal processing techniques \cite{FilteredRB-1}, which exhibit robustness issues. The lifting framework that we introduce provides a systematic solution to the ``multiplicity problem'' in randomized benchmarking by enabling clean, single-exponential decay fits even when the superoperator representation of the benchmarking group contains irreps with multiplicity.

Our protocol can be summarized simply as ``do tomography on the irreps.'' However, it is important to emphasize that our protocol is \emph{not} SPAM-limited. Thanks to the group-theoretic techniques that we employ, the signals that we extract remain robust to variations in the noise channels on the lifting operators and the SPAM effects. Just as gate tomography uses both state and measurement variations, we synthesize \emph{two} projection-like operators. One projector does not suffice because, in the case of multiplicity, we are estimating a matrix-valued quantity rather than a scalar-valued quantity.

Our techniques can be applied to novel quantum computing platforms.  Many emerging quantum technologies operate natively on high-dimensional \emph{qudits} (encompassing also bosons, fermions, and hybrid architectures) rather than qubits, which offer the prospect of tailored quantum error-correcting codes and increased flexibility for simulating physical systems.  Their natural gate sets often correspond to continuous groups, such as non-abelian Lie groups arising from the symmetries of their classical phase spaces.  These symmetries motivate novel \emph{group-covariant} quantum error-correcting codes \cite{Gross}.  Can benchmarking methods exploit this symmetry to enable accurate and efficient noise characterization?\footnote{One approach to this question is to find unitary designs for qudits \cite{Anand:2026dcp}, which provide systematic approximations to Haar-random unitaries.}  Synthetic RB enables the efficient benchmarking of groups that are both infinite and non-abelian, such as Lie groups relevant to high-dimensional quantum systems, and provides the first practical RB tools for high-dimensional spin qudits.  Our present work enables the application of synthetic RB in the non-multiplicity-free setting.

Lifted RB techniques can also be ported over to the problem of character eigenvalue estimation (CEE).\footnote{We thank Jordan Hines for discussions on this point.}  Given a superoperator that is guaranteed to be close to a target superoperator, we wish to estimate its (generally complex) eigenvalues to high precision with simple curve fitting.  CEE does so by combining the idea of character RB with phase estimation circuits.  The core of each CEE circuit consists of repetitions of a single gate.  One synthesizes an approximate projector using only varied-length sequences of that gate.  If that gate's target unitary implements a rotation by a rational angle and therefore generates a finite cyclic group, then we use the irreps of that group and the irrep decomposition of the superoperator representation of the target unitary to determine the weights in the synthetic circuit construction.  Irreps with multiplicity lead to (complex) multi-exponential functions that are hard to fit.  Our ZZ gate example illustrates the application of our techniques to CEE.

We expect that our techniques could find applications to the characterization of quantum error-correcting codes, through techniques similar to logical RB \cite{Combes}. In this context, by analogy to subspace and leakage RB, the codespace can be viewed as the computational subspace. The Hilbert space of the codespace is of smaller dimension than the Hilbert space of the data qubit register, and the representation of the codespace projector's action on density matrices will have multiplicity. We expect that our techniques could be deployed to measure the quality of a codespace projector and possibly other aspects of the code performance. This line of effort will be future work. 

Finally, we leave for future work the development of further mathematical foundations and guarantees for lifted RB, including a systematic analysis of gate-dependent noise \cite{Proctor, Wallman, Merkel, Claes}.  For instance, it would be interesting to obtain analytical estimates and error bounds for the eigenvalues of the matrices $M_\lambda(\mathcal{E})$ in perturbation theory around the zero-noise limit.  This line of work has bearing on a number of practical questions.  To what precision can we measure the average fidelity of the benchmarking group, as a function of the level of noise present in the gates?  If the gates in the lifting group are much noisier than those in the benchmarking group, does this affect the precision of the estimate?

\section*{Acknowledgements}

We thank Jordan Hines for initial collaboration on this project. JPM thanks Winton Brown and Ryan Bilotta for insightful conversations. Generative AI tools were used to assist with coding.  All calculations and text in this paper are human-generated.

\appendix

\section{Group Twirling} \label{app:twirling}

Let $\phi$ be any unitary representation of a compact group $G$ on a complex vector space $V$, let $X$ be a linear operator on $V$, and let
\begin{equation}
\Phi(X)\equiv \mathbb{E}_{g\in G}[\phi(g)X\phi(g)^\dag].
\end{equation}
We refer to $\Phi(X)$ as the $G$-twirl of $X$.  For any $g\in G$, some reindexing shows that
\begin{equation}
\phi(g)\Phi(X) = \Phi(X)\phi(g).
\label{reindexing}
\end{equation}
For convenient reference, we describe the implications of Schur's lemma for the form of $\Phi(X)$ in increasing generality (see also \cite{Gambetta, Fulton-Harris}).

\subsection{Irreducible Representation}

If $\phi$ is irreducible, then Schur's lemma implies that $\Phi(X)\propto \mathds{1}$, where the constant of proportionality can be determined by taking the trace:
\begin{equation}
\Phi(X) = \frac{\tr(X)}{\dim(V)}\mathds{1}.
\end{equation}

\subsection{Multiplicity-Free Representation}

Now suppose that $\phi$ is reducible but multiplicity-free: we write $\phi\cong \bigoplus_\lambda \phi_\lambda$ and $V\cong \bigoplus_\lambda V_\lambda$, where $\lambda$ labels inequivalent irreps.  Let $\Pi_\lambda$ denote the projector onto the support $V_\lambda$ of $\phi_\lambda$.  Using $\Pi_\lambda\phi(g)\Pi_\lambda \linebreak[1] = \linebreak[1] \phi_\lambda(g)$ and inserting projectors into \eqref{reindexing}, we have
\begin{equation}
\phi_\lambda(g)(\Pi_\lambda\Phi(X)\Pi_{\lambda'}) = (\Pi_\lambda\Phi(X)\Pi_{\lambda'})\phi_{\lambda'}(g).
\end{equation}
Thus the linear map $\Pi_\lambda\Phi(X)\Pi_{\lambda'}$ is an intertwiner for the irreps $\lambda$ and $\lambda'$.  By Schur's lemma, it satisfies
\begin{equation}
\Pi_\lambda\Phi(X)\Pi_{\lambda'}\propto \delta_{\lambda\lambda'}\Pi_\lambda.
\end{equation}
The constant of proportionality can be determined by taking the trace:
\begin{equation}
\Pi_\lambda\Phi(X)\Pi_\lambda = \frac{\tr(X\Pi_\lambda)}{\tr(\Pi_\lambda)}\Pi_\lambda.
\end{equation}
We conclude that
\begin{equation}
\Phi(X) = \mathds{1}\Phi(X)\mathds{1} = \left(\sum_\lambda \Pi_\lambda\right)\Phi(X)\left(\sum_{\lambda'} \Pi_{\lambda'}\right) = \sum_{\lambda, \lambda'} \Pi_\lambda\Phi(X)\Pi_{\lambda'} = \sum_\lambda \frac{\tr(X\Pi_\lambda)}{\tr(\Pi_\lambda)}\Pi_\lambda.
\end{equation}

\subsection{General Representation}

Finally, let $\phi$ be a completely general representation in which the irrep $\phi_\lambda$ has multiplicity $n_\lambda$:
\begin{equation}
\phi\cong \bigoplus_\lambda \phi_\lambda^{\oplus n_\lambda}, \qquad V\cong \bigoplus_\lambda \mathbb{C}^{n_\lambda}\otimes V_\lambda.
\end{equation}
We denote the copies by $\phi_{\lambda^{(i)}}$ and their corresponding projectors by $\Pi_{\lambda^{(i)}}$, where $i = 1, \ldots, n_\lambda$.  Again inserting projectors into \eqref{reindexing}, we find that
\begin{equation}
\phi_{\lambda^{(i)}}(g)(\Pi_{\lambda^{(i)}}\Phi(X)\Pi_{\lambda'^{(i')}}) = (\Pi_{\lambda^{(i)}}\Phi(X)\Pi_{\lambda'^{(i')}})\phi_{\lambda'^{(i')}}(g).
\end{equation}
Schur's lemma implies that if $\lambda = \lambda'$, then $\Pi_{\lambda^{(i)}}\Phi(X)\Pi_{\lambda'^{(i')}}$ is an isomorphism between $\phi_{\lambda^{(i)}}$ and $\phi_{\lambda^{(i')}}$.  If, in addition, $i = i'$, then $\Pi_{\lambda^{(i)}}\Phi(X)\Pi_{\lambda'^{(i')}}$ is proportional to $\Pi_{\lambda^{(i)}}$.  Concretely, for each $i$, let
\begin{equation}
\{|\lambda^{(i)}, \ell\rangle \,|\, \ell = 1, \ldots, \dim(V_\lambda)\}
\end{equation}
be an orthonormal basis for the support of $\phi_{\lambda^{(i)}}$ such that the group representation matrices for different copies $i$ of the same irrep $\lambda$ are identical.  Explicitly, we have
\begin{equation}
\phi(g) = \bigoplus_\lambda\bigoplus_{i=1}^{n_\lambda} \phi_{\lambda^{(i)}}(g) = \sum_\lambda\sum_{i=1}^{n_\lambda}\sum_{\ell, \ell' = 1}^{\dim(V_\lambda)} \phi_{\lambda, \ell, \ell'}(g)|\lambda^{(i)}, \ell\rangle\langle\lambda^{(i)}, \ell'|
\end{equation}
where $\phi_\lambda(g)$ is a unitary matrix:
\begin{equation}
\sum_{k=1}^{\dim(V_\lambda)} \phi_{\lambda, \ell, k}(g)\phi_{\lambda, \ell', k}(g)^\ast = \sum_{k=1}^{\dim(V_\lambda)} \phi_{\lambda, k, \ell}(g)\phi_{\lambda, k, \ell'}(g)^\ast = \delta_{\ell\ell'}.
\end{equation}
In addition, let
\begin{equation}
Q_{\lambda, i, i'}\equiv \sum_{\ell=1}^{\dim(V_\lambda)} |\lambda^{(i)}, \ell\rangle\langle\lambda^{(i')}, \ell|
\end{equation}
(note that $\Pi_{\lambda^{(i)}} = Q_{\lambda, i, i}$).  Then Schur's lemma implies that
\begin{equation}
\Pi_{\lambda^{(i)}}\Phi(X)\Pi_{\lambda'^{(i')}}\propto \delta_{\lambda\lambda'}Q_{\lambda, i, i'}.
\end{equation}
Using $\tr(Q_{\lambda, i, i'}Q_{\lambda, j, j'}) = \delta_{ij'}\delta_{i'j}\dim(V_\lambda)$ and writing
\begin{equation}
\tr(Q_{\lambda, i', i}\Pi_{\lambda^{(i)}}\Phi(X)\Pi_{\lambda^{(i')}}) = \tr(Q_{\lambda, i', i}\Phi(X)) = \tr(Q_{\lambda, i', i}X)
\label{tracewithQ}
\end{equation}
allows us to determine the constant of proportionality:
\begin{equation}
\Pi_{\lambda^{(i)}}\Phi(X)\Pi_{\lambda^{(i')}} = \frac{\tr(Q_{\lambda, i', i}X)}{\dim(V_\lambda)}Q_{\lambda, i, i'}.
\end{equation}
In writing the last equality of \eqref{tracewithQ}, we have used $\phi(g)^\dag Q_{\lambda, i', i}\phi(g) = Q_{\lambda, i', i}$, which relies on the fact that $\phi_{\lambda^{(i)}}(g) = \phi_\lambda(g)$ for all $i$ in our chosen basis.  We conclude that
\begin{equation}
\Phi(X) = \sum_\lambda\sum_{i, i' = 1}^{n_\lambda} \Pi_{\lambda^{(i)}}\Phi(X)\Pi_{\lambda^{(i')}} = \sum_\lambda\sum_{i, i' = 1}^{n_\lambda} \frac{\tr(Q_{\lambda, i', i}X)}{\dim(V_\lambda)}Q_{\lambda, i, i'}.
\end{equation}
Letting
\begin{equation}
[M_\lambda(X)]_{ii'}\equiv \frac{\tr(Q_{\lambda, i', i}X)}{\dim(V_\lambda)}
\end{equation}
where $[M_\lambda(X)]_{ii'}$ is the $(i, i')$ entry of the $n_\lambda\times n_\lambda$ matrix $M_\lambda(X)$, we can equivalently write
\begin{equation}
\Phi(X) = \sum_\lambda\sum_{i, i' = 1}^{n_\lambda} [M_\lambda(X)]_{ii'}Q_{\lambda, i, i'} = \bigoplus_\lambda M_\lambda(X)\otimes \mathds{1}_\lambda,
\label{phitensor}
\end{equation}
as in \cite{Claes}.

Now consider an arbitrary basis.  Any orthonormal basis can be reached from a basis of the above form by a unitary transformation.  Any change of orthonormal basis within a given irrep is described by a unitary matrix $U_{\lambda^{(i)}}$:
\begin{equation}
|\lambda^{(i)}, \ell\rangle\to \sum_{\ell'} [U_{\lambda^{(i)}}]_{\ell'\ell}|\lambda^{(i)}, \ell'\rangle
\end{equation}
(note the order of the indices; this transformation rule pertains to the basis vectors rather than the coefficients).  Writing
\begin{equation}
\phi_{\lambda^{(i)}}(g) = \sum_{\ell, \ell'} \phi_{\lambda, \ell, \ell'}(g)|\lambda^{(i)}, \ell\rangle\langle\lambda^{(i)}, \ell'|,
\end{equation}
we then have
\begin{align}
\phi_{\lambda^{(i)}}(g) &\to \sum_{k, k'} \left(\sum_{\ell, \ell'} [U_{\lambda^{(i)}}]_{k\ell}\phi_{\lambda, \ell, \ell'}(g)[U_{\lambda^{(i)}}]_{k'\ell'}^\ast\right)|\lambda^{(i)}, k\rangle\langle\lambda^{(i)}, k'| \\
&= \sum_{k, k'} [U_{\lambda^{(i)}}\phi_\lambda(g)U_{\lambda^{(i)}}^\dag]_{kk'}|\lambda^{(i)}, k\rangle\langle\lambda^{(i)}, k'|
\end{align}
and
\begin{align}
Q_{\lambda, i, i'} &\to \sum_{\ell', \ell''} \left(\sum_\ell [U_{\lambda^{(i)}}]_{\ell'\ell}[U_{\lambda^{(i')}}]_{\ell''\ell}^\ast\right)|\lambda^{(i)}, \ell'\rangle\langle\lambda^{(i')}, \ell''| \\
&= \sum_{\ell', \ell''} [U_{\lambda^{(i)}}U_{\lambda^{(i')}}^\dag]_{\ell'\ell''}|\lambda^{(i)}, \ell'\rangle\langle\lambda^{(i')}, \ell''|.
\end{align}
Therefore, in a general basis, the blocks of $\Phi(X)$ for each $\lambda$ no longer take the form of tensor products.  Regardless, there always exists a basis in which $\Phi(X)$ takes the form \eqref{phitensor}.

\section{Generalized Completeness Relation} \label{app:completeness}

The key to building synthetic gates is the generalized completeness relation \eqref{completeness}. (See \cite{Fan:2024puh} for a derivation.) For our purposes, we need a further generalization of this formula, \eqref{completenessgen}, that handles multiplicities.

To derive it, we begin with the group twirling relation for a general representation:
\begin{equation}
\mathbb{E}_{g\in G}[\phi(g)X\phi(g)^\dag] = \sum_\lambda\sum_{i, i' = 1}^{n_\lambda} \frac{\tr(Q_{\lambda, i', i}X)}{\dim(V_\lambda)}Q_{\lambda, i, i'}.
\end{equation}
Letting $\{|\alpha\rangle\}$ be an orthonormal basis for $V$, so that $\{e_{\alpha\beta}\equiv |\alpha\rangle\langle\beta|\}$ is an orthonormal basis for $\mathcal{L}(V)$, this is equivalent to
\begin{equation}
\mathbb{E}_{g\in G}[\langle\alpha|\phi(g)|\beta\rangle\langle\beta'|\phi(g)^\dag|\alpha'\rangle] = \sum_\lambda\sum_{i, i' = 1}^{n_\lambda} \frac{\langle\alpha|Q_{\lambda, i, i'}|\alpha'\rangle\langle\beta'|Q_{\lambda, i', i}|\beta\rangle}{\dim(V_\lambda)}
\end{equation}
for all $\alpha, \alpha', \beta, \beta'$.  In terms of superkets, this is
\begin{equation}
\mathbb{E}_{g\in G}[\llangle e_{\alpha\beta}|\phi(g)\rrangle\llangle\phi(g)|e_{\alpha'\beta'}\rrangle] = \sum_\lambda\sum_{i, i' = 1}^{n_\lambda} \frac{\llangle e_{\alpha\beta}|Q_{\lambda, i, i'}e_{\alpha'\beta'}Q_{\lambda, i', i}\rrangle}{\dim(V_\lambda)}.
\end{equation}
By linearity, this is equivalent to
\begin{equation}
\mathbb{E}_{g\in G}[|\phi(g)\rrangle\llangle\phi(g)|X\rrangle] = \sum_\lambda\sum_{i, i' = 1}^{n_\lambda} \frac{|Q_{\lambda, i, i'}XQ_{\lambda, i', i}\rrangle}{\dim(V_\lambda)}
\end{equation}
for all operators $X$.  We can rewrite this as
\begin{equation}
\mathbb{E}_{g\in G}[\llangle\phi(g)|X\rrangle\phi(g)] = \sum_\lambda\sum_{i, i' = 1}^{n_\lambda} \frac{Q_{\lambda, i, i'}XQ_{\lambda, i', i}}{\dim(V_\lambda)}.
\end{equation}
If $X\equiv X_\lambda$ acts within the $\lambda$-isotypic component, then this gives
\begin{equation}
\dim(V_\lambda)\mathbb{E}_{g\in G}[\llangle\phi(g)|X_\lambda\rrangle\phi(g)] = \sum_{i, i' = 1}^{n_\lambda} Q_{\lambda, i, i'}X_\lambda Q_{\lambda, i', i}.
\end{equation}
In particular, if $X = \Pi_{\lambda^{(j)}} = Q_{\lambda, j, j}$ for any fixed $j$, then this gives
\begin{equation}
\dim(V_\lambda)\mathbb{E}_{g\in G}[\chi_\lambda(g)^\ast\phi(g)] = \sum_{i, i' = 1}^{n_\lambda} Q_{\lambda, i, i'}Q_{\lambda, j, j}Q_{\lambda, i', i} = \sum_{i=1}^{n_\lambda} Q_{\lambda, i, j}Q_{\lambda, j, j}Q_{\lambda, j, i} = \sum_{i=1}^{n_\lambda} Q_{\lambda, i, i} = \Pi_\lambda.
\end{equation}
More generally, even if $X$ acts within a single copy of the irrep $\phi_\lambda$, the synthesized operator will not act only within that copy because the $Q$'s ``connect'' $X$ to other copies within the $\lambda$-isotypic component.

Indeed, let $X = |\lambda^{(j)}, \ell\rangle\langle\lambda^{(j')}, \ell'|$.  Then
\begin{equation}
Q_{\lambda, i, i'}XQ_{\lambda, i', i} = \delta_{i'j}\delta_{i'j'}|\lambda^{(i)}, \ell\rangle\langle\lambda^{(i)}, \ell'|
\end{equation}
and hence
\begin{equation}
\sum_{i, i' = 1}^{n_\lambda} Q_{\lambda, i, i'}XQ_{\lambda, i', i} = \delta_{jj'}\sum_{i=1}^{n_\lambda} |\lambda^{(i)}, \ell\rangle\langle\lambda^{(i)}, \ell'|.
\end{equation}
So if $X$ is a matrix unit within the $\lambda$-isotypic component, then for the synthesized operator to be nontrivial, $X$ must be a matrix unit within a single copy.

\section{Benchmarking the Trivial Representation} \label{app:trivial}

Here, we comment on a subtlety involving the trivial representation.  No lifting group (even the full unitary group) has a superoperator representation with a nontrivial irrep that contains the entire isotypic component of the trivial representation in the superoperator representation of the benchmarking group.  However, this seeming difficulty is surmounted by the fact that one of the RB quality parameters associated with the trivial representation is automatically 1.  This fact obviates the need to perform full tomography on the matrix $M_0(\mathcal{E})$.

To see this, suppose we choose an orthonormal basis $\{O_i\}$ of operators on $\mathbb{C}^d$ such that $O_0 = \mathds{1}/\sqrt{d}$.  The operators $O_{i>1}$ are therefore traceless.  A trace-preserving superoperator $\Lambda$ satisfies
\begin{equation}
\llangle O_0|\Lambda|O_i\rrangle = \tr(O_0^\dag\Lambda(O_i)) = \delta_{0i}\tr(O_0^\dag\Lambda(O_0)) = \delta_{0i},
\end{equation}
whereas a unital superoperator $\Lambda$ satisfies
\begin{equation}
\llangle O_i|\Lambda|O_0\rrangle = \tr(O_i^\dag\Lambda(O_0)) = \tr(O_i^\dag O_0) = \delta_{i0}.
\end{equation}
A unitary channel (in particular, any group element superoperator $\mathcal{G}$) is both trace-preserving and unital.

The superoperator (conjugation) representation of any benchmarking group always contains a copy of the trivial representation (call it $\phi_0$) spanned by the identity operator because unitary channels are unital.  If it occurs without multiplicity, then we have
\begin{equation}
\Tr(\Pi_0\mathcal{E}) = \llangle O_0|\mathcal{E}|O_0\rrangle = \tr(O_0^\dag\mathcal{E}(O_0)) = \frac{1}{d}\tr(\mathcal{E}(\mathds{1})) = 1
\end{equation}
because the noise channel $\mathcal{E}$ is trace-preserving.  If it occurs with some multiplicity $n_0$, then let us denote the copies by
\begin{equation}
\phi_{0^{(i)}} = \operatorname{span}\{O_{0^{(i)}}\}
\end{equation}
for $i = 1, \ldots, n_0$, where $\phi_{0^{(1)}} = \phi_0$ and $O_{0^{(1)}} = O_0$.  We have
\begin{equation}
[M_0(\mathcal{E})]_{ii'} = \Tr(Q_{0, i', i}\mathcal{E}) = \llangle O_{0^{(i)}}|\mathcal{E}|O_{0^{(i')}}\rrangle \implies [M_0(\mathcal{E})]_{1i} = \delta_{1i}
\end{equation}
because $\mathcal{E}$ is trace-preserving.  Thus the matrix $M_0(\mathcal{E})$ takes the block triangular form
\begin{equation}
M_0(\mathcal{E}) = \left[\begin{array}{cc}
1 & \mathbf{0}_{1\times (n_0 - 1)} \\
\ast_{(n_0 - 1)\times 1} & \widetilde{M}_{(n_0 - 1)\times (n_0 - 1)}
\end{array}\right]
\end{equation}
and, in particular, has a left eigenvector $(1, 0, \ldots, 0)$ with eigenvalue 1.  The other eigenvalues of $M_0(\mathcal{E})$ are simply those of $\widetilde{M}$, irrespective of the entries denoted by $\ast$.

Since one eigenvalue of $M_0(\mathcal{E})$ is always 1, to determine the other eigenvalues, it suffices to perform tomography on the matrix $\widetilde{M}$ by synthesizing operators of the form $Q_{0, i, i'}$ with $i, i' > 1$.  This can be done using group element superoperators.

\section{Choosing the Lifting Group} \label{app:lifting}

Here, we discuss some general strategies for choosing the lifting group.  Aside from using the full unitary group, we can attempt to construct a lifting group by supplementing the benchmarking group with a finite set of experimentally relevant gates (which may nonetheless result in an infinite lifting group).  These additional gates may be targeted to the isotypic component of one irrep\footnote{An ad hoc approach is to choose additional gates that mix up states within the isotypic component of a target irrep but that act trivially within all other irreps, assuming that such gates exist.} or to all isotypic components simultaneously; for simplicity, we consider the latter, as it reduces casework.

For a completely reducible representation $V$ of a group $G$ that decomposes as $\bigoplus_\lambda (\mathbb{C}^{n_\lambda} \linebreak[1] \otimes \linebreak[1] V_\lambda)$, Schur's lemma implies that any operator in the commutant algebra $\operatorname{End}_G(V)$ acts as the identity on the irreps $V_\lambda$ and only acts nontrivially on the multiplicity spaces $\mathbb{C}^{n_\lambda}$.  That is,
\begin{equation}
\operatorname{End}_G(V)\cong \bigoplus_\lambda \operatorname{End}(\mathbb{C}^{n_\lambda})\cong \bigoplus_\lambda M_{n_\lambda}(\mathbb{C}).
\end{equation}
Suppose we are given the generators of a finite group as finite-dimensional matrices acting on a vector space $V$.  Assuming we do not already know the irrep decomposition and irreducible characters, a quick way to determine the irreps that occur inside the given representation on $V$ is to diagonalize a generic element of the commutant.  If no irreps occur with multiplicity, then the eigenspaces yield the irreps.  With multiplicity, the eigenspaces again determine the irreps, but non-canonically inside each isotypic component.

More precisely, suppose $G$ is a finite group represented on a finite-dimensional complex vector space $V$.  For a generic element $T$ of the commutant $\operatorname{End}_G(V)$, the eigenspaces of $T$ are irreducible $G$-subrepresentations of $V$.  If the representation is multiplicity-free, then these eigenspaces are the canonical irreducible summands.  If an irrep occurs with multiplicity, then the eigenspaces select a noncanonical decomposition of its isotypic component into irreducible copies.  Indeed, while the isotypic decomposition of $V$ is canonical, choosing a decomposition of an isotypic component into irreps is tantamount to choosing a basis for the corresponding multiplicity space.

If one adjoins a generic invertible element of the commutant to the original generating set, then one obtains a representation of a new group whose irreps are multiplicity-free.  Specifically, a generic invertible element of the commutant produces a multiplicity-free refinement of the original representation whose irreducible summands are individual copies of the original irreps.  An element $T$ of the commutant takes the form
\begin{equation}
T = \bigoplus_\lambda (T_\lambda\otimes I_{V_\lambda}).
\end{equation}
For generic $T$, each $T_\lambda$ has a nondegenerate spectrum.  Hence
\begin{equation}
V\cong \bigoplus_{\lambda, j} \left(\ell_{\lambda, j}\otimes V_\lambda\right),
\end{equation}
where $\ell_{\lambda, j}$ are the one-dimensional eigenspaces of $T_\lambda$.  Each summand $\ell_{\lambda, j}\otimes V_\lambda$ is invariant under both $G$ and $T$, and it is irreducible for the enlarged action.  Moreover, the summands are non-isomorphic as representations of the enlarged group because the extra generator $T$ distinguishes the different copies of $V_\lambda$ by assigning them different eigenvalues.  Hence the enlarged representation is generically multiplicity-free.\footnote{If $G\cap \langle T\rangle = \{I\}$, then the enlarged matrix group $\langle G, T\rangle$ is a direct product $G\times \langle T\rangle$.}

If one adjoins \emph{two} generic and non-commuting (and invertible) elements of the commutant $\operatorname{End}_G(V)$ to the original generating set, then the irreducible subspaces for the enlarged action are exactly the original isotypic components $\mathbb{C}^{n_\lambda}\otimes V_\lambda$.  To see this, write $T, S\in \operatorname{End}_G(V)$ as
\begin{equation}
T = \bigoplus_\lambda (T_\lambda\otimes I_{V_\lambda}), \qquad S = \bigoplus_\lambda (S_\lambda\otimes I_{V_\lambda}).
\end{equation}
Two generic non-commuting matrices $T_\lambda$ and $S_\lambda$ have no common proper invariant subspace.  Equivalently, the algebra they generate is the full matrix algebra:
\begin{equation}
\mathbb{C}\langle T_\lambda, S_\lambda\rangle = \operatorname{End}(\mathbb{C}^{n_\lambda}).
\end{equation}
This is a consequence of Burnside's theorem, which states that a matrix algebra over $\mathbb{C}$ acts irreducibly precisely when it is the full matrix algebra.  Therefore, on the $\lambda$-isotypic component, the algebra generated by the original representation together with $T$ and $S$ is
\begin{equation}
\operatorname{End}(\mathbb{C}^{n_\lambda})\otimes \operatorname{End}(V_\lambda)\cong \operatorname{End}(\mathbb{C}^{n_\lambda}\otimes V_\lambda).
\end{equation}
Since the full endomorphism algebra acts irreducibly, $\mathbb{C}^{n_\lambda}\otimes V_\lambda$ is irreducible for the enlarged group action.\footnote{As before, if $G\cap \langle T, S\rangle = \{I\}$, then the enlarged group $\langle G, T, S\rangle$ is a direct product $G\times \langle T, S\rangle$.}  This observation yields a general strategy for choosing the lifting group, as long as we work with the \emph{superoperator} representations of the relevant gates.

We let $G_x$, $G_y$, $G_{zz}$ denote the superoperator representations of $M_x$, $M_y$, $M_{zz}$ as $16\times 16$ matrices, respectively.  A convenient operator basis is the orthonormal Pauli basis
\begin{equation}
\frac{1}{2}(II, IX, IY, IZ, XI, XX, XY, XZ, YI, YX, YY, YZ, ZI, ZX, ZY, ZZ),
\end{equation}
in which the superoperator representation is the Pauli transfer matrix (PTM) representation.

\subsection{Example: ZZ Gate}

In this example, the benchmarking group is a subgroup of the two-qubit Clifford group, which already suffices as a lifting group.  Nonetheless, it is instructive to see how our general strategy for choosing a lifting group might apply here.

As remarked above, it is not enough to construct the lifting group by adjoining unitary gates that belong to the commutant of the benchmarking group.  We must adjoin \emph{superoperators} that belong to the commutant of the \emph{superoperator representation} of the benchmarking group.  However, not every superoperator arises as the conjugation representation of a unitary gate, so our direct strategy is not guaranteed to work for every irrep.  Nonetheless, it is still useful to consider the commutant algebra (or the unitary commutant group) of the benchmarking group because the irreps in its superoperator representation are guaranteed to be no larger than the isotypic components of the irreps in the superoperator representation of the benchmarking group.  For sample efficiency, it is beneficial to minimize the sizes of the irreps that occur in the superoperator representation of the lifting group.

Recall that $\langle M_{zz}\rangle = \mathbb{Z}_8$ and $\langle G_{zz}\rangle = \mathbb{Z}_4$.  The irrep decomposition of the conjugation representation of $\langle M_{zz}\rangle$ on the space of 4-by-4 complex matrices $\operatorname{End}(\mathbb{C}^4)$ follows from diagonalizing $G_{zz}$: the eigenvalues are 1 with multiplicity 8 and $\pm i$ with multiplicity 4 each.

The commutant of $M_{zz}$ consists of all block matrices of the form
\begin{equation}
\left[\begin{array}{cccc} a_1 & 0 & 0 & a_2 \\ 0 & b_1 & b_2 & 0 \\ 0 & b_3 & b_4 & 0 \\ a_3 & 0 & 0 & a_4 \end{array}\right],
\end{equation}
which contains the group $S(U(2)\times U(2))$ of parity-preserving unitaries on two qubits.  For simplicity, we ignore the restriction on the overall phase and consider $U(2)\times U(2)$.  We are concerned with the 16-dimensional conjugation representation of $U(2)\times U(2)$ on $\operatorname{End}(\mathbb{C}^4)$.  We can decompose this representation into irreps as follows.  First,
\begin{equation}
V\equiv \mathbb{C}^4 = V_\text{even}\oplus V_\text{odd},
\end{equation}
where the two two-dimensional subspaces have definite parity.  Our group is then $U(V_\text{even})\times U(V_\text{odd})$.  Since $\operatorname{End}(V) = V\otimes V^\ast$, we have
\begin{equation}
\operatorname{End}(V) = \operatorname{End}(V_\text{even})\oplus \operatorname{End}(V_\text{odd})\oplus \operatorname{Hom}(V_\text{odd}, V_\text{even})\oplus \operatorname{Hom}(V_\text{even}, V_\text{odd}).
\end{equation}
Each $\operatorname{End}(V_\star)$ summand decomposes as
\begin{equation}
\operatorname{End}(V_\star) = \text{triv}\oplus \operatorname{Ad}_\star,
\end{equation}
where $\operatorname{Ad}_\star$ is three-dimensional.  Let $S_\text{even}$, $S_\text{odd}$ denote the standard two-dimensional representations of the two $U(2)$ factors.  Then
\begin{equation}
\operatorname{Hom}(V_\text{odd}, V_\text{even})\cong S_\text{even}\otimes S_\text{odd}^\ast, \qquad \operatorname{Hom}(V_\text{even}, V_\text{odd})\cong S_\text{even}^\ast\otimes S_\text{odd}.
\end{equation}
Therefore,
\begin{equation}
\operatorname{End}(V)\cong \underbrace{2(\text{triv})\oplus \operatorname{Ad}_\text{even}\oplus \operatorname{Ad}_\text{odd}}_{\text{$\phi_0$-isotypic}}\oplus \underbrace{(S_\text{even}\otimes S_\text{odd}^\ast)}_{\text{$\phi_+$-isotypic}}\oplus \underbrace{(S_\text{even}^\ast\otimes S_\text{odd})}_{\text{$\phi_-$-isotypic}}.
\label{endVdecomposition}
\end{equation}
There are six irreducible summands, with the trivial representation occurring with multiplicity 2.  Importantly, while the multiplication representation of $U(2)\times U(2)$ on $\mathbb{C}^4$ generates the commutant algebra of the multiplication representation of $\langle M_{zz}\rangle$ on $\mathbb{C}^4$, the conjugation representation of $U(2)\times U(2)$ on $\operatorname{End}(\mathbb{C}^4)$ does \emph{not} generate the commutant algebra of the conjugation representation of $\langle M_{zz}\rangle$ on $\operatorname{End}(\mathbb{C}^4)$.  Therefore, we conclude that $U(2)\times U(2)$ \emph{preserves} the isotypic components of the $\langle M_{zz}\rangle$ representation on $\operatorname{End}(\mathbb{C}^4)$, but it may decompose those isotypic components further, as indicated by the underbraces in \eqref{endVdecomposition}.  Indeed, the $\phi_\pm$-isotypic components are irreducible for $U(2)\times U(2)$, but the $\phi_0$-isotypic component is not.  The full commutant algebra of $G_{zz}$ is
\begin{equation}
\operatorname{End}(V_0)\oplus \operatorname{End}(V_+)\oplus \operatorname{End}(V_-)\cong M_8(\mathbb{C})\oplus M_4(\mathbb{C})\oplus M_4(\mathbb{C}),
\end{equation}
which has dimension 96.

In summary, $U(2)\times U(2)$ is a subgroup of $U(4)$ that can be used as a lifting group for the $\phi_+$ or $\phi_-$ irreps of the benchmarking group $\langle M_{zz}\rangle$.  Note that $U(2)\times U(2)$ is not much more practical to implement than the full unitary group $U(4)$, as unitaries acting within definite-parity sectors are generically two-qubit entangling gates.  Moreover, it is not a finite group.  However, randomly sampling from an infinite group is precisely the situation where synthetic RB techniques \cite{Fan:2024puh} are most useful.

\subsection{Example: MS Gates} \label{app:msgates}

The MS benchmarking group $\langle M_x, M_y\rangle$ has order 192, while its projective version $\langle G_x, G_y\rangle$ has order 96.  The commutant algebra of $M_x$ and $M_y$ consists of all matrices of the form
\begin{equation}
\left[\begin{array}{cccc}
a & 0 & 0 & 0 \\
0 & b & c & 0 \\
0 & c & b & 0 \\
0 & 0 & 0 & a
\end{array}\right]
\end{equation}
for $a, b, c\in \mathbb{C}$.  This algebra is commutative, so our general strategy does not apply here.

Note that the MS benchmarking group is a subgroup of $U(2)_\text{even}\times U(2)_\text{odd}$, so the irreps of the former must lie inside the irreps of the latter.  Indeed, using the notation of \eqref{orderedbasis1} and \eqref{orderedbasis2}, we have
\begin{align}
\operatorname{span}\{\rho_1^{(1)}, \rho_2^{(1)}\} &= \operatorname{span}\left\{\frac{1}{4}(IX + XI + iYZ + iZY), \frac{1}{4}(XZ + ZX + iIY + iYI)\right\}, \\
\operatorname{span}\{\rho_3^{(1)}, \rho_4^{(1)}\} &= \operatorname{span}\left\{\frac{1}{4}(IX - XI + iYZ - iZY), \frac{1}{4}(XZ - ZX + iIY - iYI)\right\}, \\
\operatorname{span}\{\rho_1^{(2)}, \rho_2^{(2)}\} &= \operatorname{span}\left\{\frac{1}{4}(IX - XI - iYZ + iZY), \frac{1}{4}(XZ - ZX - iIY + iYI)\right\}, \\
\operatorname{span}\{\rho_3^{(2)}, \rho_4^{(2)}\} &= \operatorname{span}\left\{\frac{1}{4}(IX + XI - iYZ - iZY), \frac{1}{4}(XZ + ZX - iIY - iYI)\right\}.
\end{align}
Therefore, we see that
\begin{align}
\operatorname{span}\{\rho_1^{(1)}, \rho_2^{(1)}\} + \operatorname{span}\{\rho_1^{(2)}, \rho_2^{(2)}\} &= \text{$\phi_+$-isotypic component}, \\
\operatorname{span}\{\rho_3^{(1)}, \rho_4^{(1)}\} + \operatorname{span}\{\rho_3^{(2)}, \rho_4^{(2)}\} &= \text{$\phi_-$-isotypic component},
\end{align}
where the $\phi_\pm$-isotypic components of the ZZ benchmarking group are irreps of $U(2)\times U(2)$.  This implies that $U(2)\times U(2)$ cannot be used as a lifting group for the complex-conjugate irreps $\phi_{b, 1}$ and $\phi_{b, 2}$ of the MS benchmarking group.

In this example, there are two physically motivated options for the lifting group, namely the group generated by the two-qubit entangling gates $M_x$ and $M_y$ along with either \cite{Sutherland:2025shv}:
\begin{enumerate}
\item single-qubit $X(\pi/2)$, $Y(\pi/2)$, and a discrete set of $Z$-rotations, or
\item single-qubit Clifford gates,
\end{enumerate}
where the single-qubit $P(\pi/2)$ gates are
\begin{equation}
e^{i(\pi/4)P\otimes I} = e^{i(\pi/4)P}\otimes I, \qquad e^{i(\pi/4)I\otimes P} = I\otimes e^{i(\pi/4)P}.
\end{equation}
These two options are essentially the same because the gates $e^{i(\pi/4)X}$ and $e^{i(\pi/4)Y}$ themselves generate the single-qubit Clifford group.\footnote{The single-qubit gates $e^{i(\pi/4)X}$, $e^{i(\pi/4)Y}$, $e^{i(\pi/4)Z}$ generate the binary octahedral group $2O$, which is projectively the single-qubit Clifford group $S_4$.  In fact, the gates $e^{i(\pi/4)X}$ and $e^{i(\pi/4)Y}$ themselves generate the gate $e^{i(\pi/4)Z}$, as $e^{i(\pi/4)Z} = e^{i(\pi/4)X}e^{-i(\pi/4)Y}e^{-i(\pi/4)X}$.}  Moreover, the group generated by $M_x$, $M_y$, and single-qubit Clifford gates is (projectively) dense in $SU(4)$, hence universal.  To see this, we argue as follows.
\begin{itemize}
\item First, local Clifford gates together with $M_x$ generate the full two-qubit Clifford group.  As $M_x = e^{i(\pi/4)X\otimes X}$ and $HXH = Z$, we have
\begin{equation}
e^{i(\pi/4)Z\otimes Z} = (H\otimes H)M_x(H\otimes H).
\end{equation}
We also have
\begin{equation}
\text{CZ} = e^{-i\pi/4}(S\otimes S)e^{i(\pi/4)Z\otimes Z}, \qquad \text{CNOT} = (I\otimes H)\text{CZ}(I\otimes H).
\end{equation}
Putting these together gives
\begin{equation}
\text{CNOT} = e^{-i\pi/4}(SH\otimes HSH)M_x(H\otimes I).
\end{equation}
Hence $H$, $S$, and $M_x$ generate CNOT; $H$, $S$, and CNOT then generate the full two-qubit Clifford group.
\item Second, one can check explicitly that $M_y$ is a nontrivial non-Clifford gate.  Therefore, the two-qubit Clifford group together with $M_y$ form a universal two-qubit gate set.
\end{itemize}
We conclude that the physically motivated lifting group is simply $SU(4)$ itself.

Suppose instead that we consider the group generated by $M_x$ and $M_y$ along with single-qubit $\pi/2$ rotations of a single type ($X$, $Y$, or $Z$).  It turns out that using either $X$ or $Y$ again leads to all of $SU(4)$.  The key fact is that
\begin{equation}
M_y = e^{i(\pi/4)Q\otimes Q}, \qquad Q\equiv \frac{X + Y}{\sqrt{2}},
\end{equation}
where $Q$ satisfies $QXQ = Y$.  This implies that $M_y$ can be used to conjugate local $X$-rotations to local $Y$-rotations, so including either local $X$- or $Y$-rotations automatically generates all local Clifford gates (and therefore all Clifford gates via $M_x$, and therefore all gates via $M_y$).  On the other hand, using $Z$ leads to a 1152-element finite group that is a subgroup of $U(2)\times U(2)$ acting within parity sectors and that, therefore, cannot be used as a lifting group for the MS benchmarking group (in fact, it is a subgroup of the finite group $2O\times 2O$).\footnote{These statements can be verified in GAP \cite{GAP4}.}  We do not attempt to construct a finite lifting group for the MS gate example.

\subsection{Non-Example: Spin Qudits}

A completely different system is a monolithic spin-$j$ qudit, whose Hilbert space $\mathscr{H}$ carries an action of the angular momentum generators $J_x, J_y, J_z$ \cite{Fan:2024puh}.  Suppose the benchmarking group were the $U(1)$ subgroup of $SU(2)$ generated by $J_z$ alone.  Its $(2j + 1)$-dimensional rep\-re\-sen\-ta\-tion on $\mathscr{H}$ is multiplicity-free and decomposes into inequivalent 1D irreps spanned by the $J_z$ eigenstates $|m\rangle$ for each $m = -j, \ldots, j$.  However, its $(2j + 1)^2$-dimensional superoperator representation on $\mathcal{L}(\mathscr{H})$ has multiplicity.  The $U(1)$ irrep labeled by $q$ (where $q = -2j, \ldots, 2j$) occurs with multiplicity $2j + 1 - |q|$, and its isotypic component is spanned by
\begin{equation}
|m + q\rangle\langle m|,
\label{u1irreps}
\end{equation}
where $m$ ranges from $-j + \max(-q, 0)$ to $j - \max(q, 0)$.  As a check, we have
\begin{equation}
\sum_{q=-2j}^{2j} (2j + 1 - |q|) = (2j + 1)^2.
\end{equation}
On the other hand, the superoperator representation of $SU(2)$ on $\mathcal{L}(\mathscr{H})$ is multiplicity-free and decomposes into $2j + 1$ irreps labeled by $k = 0, 1, \ldots, 2j$, where each irrep has dimension $2k + 1$.  Irrep $k$ is spanned by orthonormal spherical tensor operators of rank $k$:
\begin{equation}
T^{(k)}_q = \sqrt{\frac{2k + 1}{2j + 1}}\sum_{m = -j}^j C^{j, m + q}_{j, m; k, q}|m + q\rangle\langle m|, \qquad \tr((T^{(k)}_q)^\dag T^{(k')}_{q'}) = \delta_{kk'}\delta_{qq'},
\end{equation}
where $q = -k, \ldots, k$.  Thus the $SU(2)$ irrep labeled by $k$ contains $2k + 1$ $U(1)$ irreps labeled by $q$, where $q = -k, \ldots, k$.  However, these $U(1)$ irreps correspond to a different basis for the $q$-isotypic component than in \eqref{u1irreps}.  We see that the $U(1)$ irrep labeled by $q$ occurs without multiplicity inside each of the $SU(2)$ irreps labeled by $k = |q|, |q| + 1, \ldots, 2j$.  This means that the isotypic components of the $U(1)$ irreps are spread out ``as much as possible'' between distinct $SU(2)$ irreps.

This embedding of $G = U(1)\subset SU(2)$ is not ideal for the application of our technique.  For our general strategy to work, we must choose $F$ such that its superoperator representation contains a multiplicity-free irrep that contains the entire isotypic component of a target irrep of $G$.

Continuing with this non-example, suppose $SU(2)$ were the benchmarking group.  As noted in \cite{Fan:2024puh}, for sufficiently large $j$, no polyhedral subgroup of $SU(2)$ can be used as a character group for any irrep of $SU(2)$.  Moreover, the situations where a finite subgroup $H$ of $SU(2)$ can be used as the character group for the benchmarking group $SU(2)$ are precisely the situations where $SU(2)$ can be used as the lifting group for the benchmarking group $H$.  This raises a possible obstruction to performing logical RB of large spin codes \cite{Gross}: there are no examples with multiplicity where we can solve the multiplicity problem by lifting to $SU(2)$.

\begingroup
\raggedright
\bibliographystyle{quantum}
\bibliography{refs}
\endgroup

\end{document}